\pdfoutput=1
\documentclass[cernpreprint, atlasdraft=false, UKenglish, texmf, orcidlogo]{atlasdoc}

\usepackage[subfigure=true, backend=biber]{atlaspackage}
\usepackage{atlasbiblatex}

\usepackage{atlasphysics}
\graphicspath{{logos/}{figures/}}

\usepackage{ANA-TAUP-2020-03-PAPER-defs}

\usepackage{multirow}
\AtlasTitle{Reconstruction and identification of pairs of collimated $\tau$-leptons decaying hadronically using $\sqrt{s}=13$ TeV $pp$ collision data with the ATLAS detector}

\AtlasAbstract{%
This paper describes an algorithm for reconstructing and identifying a highly collimated hadronically decaying $\tau$-lepton pair with low transverse momentum. When two $\tau$-leptons are highly collimated, their visible decay products might overlap, degrading the reconstruction performance for each of the $\tau$-leptons.
A dedicated treatment attempting to tag the $\tau$-lepton pair as a single object is required.
The reconstruction algorithm is based on a large radius jet and its associated two leading subjets, and the identification uses a boosted decision tree to discriminate between signatures from $\tau^+\tau^-$ systems and those arising from QCD jets.
The efficiency of the identification algorithm is measured in $Z\gamma$ events using proton--proton collision data at $\sqrt{s}=13$ TeV collected
by the ATLAS experiment at the Large Hadron Collider between 2015 and 2018, corresponding to an integrated luminosity of $139\,\mbox{fb}^{-1}$.
The resulting data-to-simulation scale factors are close to unity with uncertainties ranging from 26\% to 37\%.
}

\AtlasRefCode{TAUP-2020-03}

\PreprintIdNumber{CERN-EP-2024-267}

\arXivId{2411.09357}

\AtlasJournal{Eur.\ Phys.\ J.\ C}
\AtlasJournalRef{Eur. Phys. J. C 85 (2025) 561}
\AtlasDOI{10.1140/epjc/s10052-025-14075-3}

\hypersetup{pdftitle={ATLAS document},pdfauthor={The ATLAS Collaboration}}

\begin{document}

\maketitle

\tableofcontents

%
%

%
\section{Introduction}
The $\tau$-lepton has a mass of $1.77$~\GeV and a lifetime of about $2.9\times 10^{-13}$~s~\cite{ParticleDataGroup:2024cfk}. It is the heaviest lepton in the Standard Model (SM), and the only lepton that can decay into hadrons, with a branching ratio to hadronic final states (\tauhad) of approximately 65\%, and to each of the light leptons of approximately 17\%.
Out of the hadronic decay modes, approximately 77\% involve one charged hadron and 22\% involve three charged hadrons. The visible part of \tauhad-lepton decays, \tauhadvis, is defined as the vectorial sum of the decay products' four-momenta, excluding neutrinos.\footnote{The \textit{vis} subscript refers to the quantities involving visible decay products.}
The reconstruction and identification of \tauhadvis is essential to many SM measurements and searches for physics beyond the SM (BSM)~\cite{ATL-PHYS-PUB-2022-044}.

In the ATLAS experiment~\cite{PERF-2007-01}, the \tauhadvis candidates are reconstructed from anti-$\kt$ jets~\cite{Cacciari:2008gp,Fastjet} with a radius parameter of 0.4, built from locally
calibrated topological clusters~\cite{PERF-2014-07}. As a result, a problem emerges when a pair of \tauhad candidates originates from the decay of a highly Lorentz-boosted parent particle (boosted di-$\tau$ system). In this scenario, the \tauhadvis pair may become too collimated to be individually resolved using standard reconstruction techniques.
When the angular distance between two \tauhadvis is smaller than the anti-$\kt$ radius parameter, the seed jet constituents may contain deposits from both of the \tauhadvis candidates resulting in a wrong grouping of these constituents by the anti-$\kt$ algorithm -- possibly in conjunction with mis-association of tracks to the correct jet -- and manifesting as a merging of the two \tauhadvis signatures into a single seed jet. It can lead to a situation in which either one or both of the \tauhadvis candidates are not correctly reconstructed with the proper track multiplicity, making it difficult to reconstruct the di-$\tau$ system from individually resolved \tauhadvis. Furthermore, even when the two \tauhadvis are reconstructed, a partial overlap of the jets can introduce problems in the identification step, possibly leading to signatures that bear more resemblance to background. These challenges require dedicated reconstruction and identification algorithms targeting boosted di-$\tau$ systems, referred to as the di-$\tau$ tagger, to recover the sensitivity to this topology.

Generally, in a two body decay, the angular distance between the decay products is approximately proportional to the ratio of the parent mass to its transverse momentum (\pt).
A high-\pt \tauhadvis pair tagger has already been developed by the ATLAS Collaboration~\cite{HDBS-2019-22}. It was used for a heavy, narrow, scalar resonance search in the high mass regime of 1--3~\TeV, decaying into a pair of Higgs bosons, where one Higgs boson decays into a $b\bar{b}$ pair and the other one into a $\tau^{+}_{\text{had}}\tau^{-}_{\text{had}}$ pair.

However, the case of lower-\pt collimated di-$\tau$ objects, relevant to BSM searches in the low-mass regime, has not yet been investigated~\cite{M_hlleitner_2017}. Several models predict the existence of light resonances~\cite{casolino2015probinglightcpoddscalar,Chang_2018} with masses smaller than half of the Higgs boson mass, produced either directly or through decays of SM particles -- most commonly via the decay of the SM Higgs boson into a pair of (pseudo)scalars~\cite{Chang:2008cw,Silveira:1985rk,Profumo:2007wc,Ipek:2014gua,Blinov:2015sna,Craig:2015pha,Bauer_2017,Derm_ek_2005,Curtin:2015fna}.
The di-$\tau$ tagging method for such final states is hence crucial for increasing the sensitivities to these signatures. %

The method presented in Ref.~\cite{HDBS-2019-22} is not directly applicable to the decays of light parent particles, as it targets a scenario in which the boosted regime corresponds to parent \pt values above approximately 300~\GeV. It uses seed jets with relatively high \pt thresholds, does not involve a dedicated energy-scale correction, and includes an identification algorithm trained to discriminate against high-\pt jet backgrounds. This method is thus unsuitable for the  case of low-mass BSM searches
and must be adapted to target signatures from the decay of light resonances with \pt smaller than 300~\GeV.
In this paper, the reconstruction and the identification of hadronically decaying collimated di-$\tau$ systems at low \pt are described and their expected performance is assessed in simulation.
The measurement of the di-$\tau$ tagger efficiency in data is presented, entailing the extraction of scale factors (SF), which account for differences in the identification efficiency between simulation and data.
The SFs are derived in a region enriched in properly identified di-$\tau$ objects, using the SM process $Z\gamma$, where the $Z$ boson decays into a boosted di-$\tau$.
The reconstruction efficiency measurement in data is beyond the scope of this paper. Additionally, the techniques presented in this paper are not applicable to semi-leptonic decays of di-$\tau$ systems, which are handled by other dedicated algorithms~\cite{TAUP-2023-02}.

After describing the ATLAS detector in Section~\ref{sec:atlasdetector} and the data and Monte Carlo simulated samples in Section~\ref{sec:mcsample}, Section~\ref{sec:reconstruction} presents the reconstruction of standard physics objects. Section~\ref{sec:ditaureconstruction} introduces the reconstruction, energy-scale calibration and identification of the boosted di-$\tau$, and Section~\ref{sec:validation} presents the SF measurement in $Z\gamma$ events. Finally, conclusions are given in Section~\ref{sec:conclusion}.

%

%
%
%
%
%
%

%
%
%
%
%
%
%

%
%
%

%
%
%
%
%
%
%
%

%
%
%
%
%


%
\newcommand{\AtlasCoordFootnote}{%
ATLAS uses a right-handed coordinate system with its origin at the nominal interaction point (IP) in the centre of the detector and the \(z\)-axis along the beam pipe. The \(x\)-axis points from the IP to the centre of the LHC ring, and the \(y\)-axis points upwards. Polar coordinates \((r,\phi)\) are used in the transverse plane, \(\phi\) being the azimuthal angle around the \(z\)-axis. The pseudorapidity is defined in terms of the polar angle \(\theta\) as \(\eta = -\ln \tan(\theta/2)\) and is equal to the rapidity $y=\frac{1}{2}\ln\left(\frac{E+p_{z}}{E-p_{z}}\right)$ in the relativistic limit. Angular distance is measured in units of \(\Delta R \equiv \sqrt{(\Delta y)^{2} + (\Delta\phi)^{2}}\).}

\section{The ATLAS detector}
\label{sec:atlasdetector}
The ATLAS detector at the LHC covers nearly the entire solid angle around the collision point.\footnote{\AtlasCoordFootnote}
It consists of an inner tracking detector surrounded by a thin superconducting solenoid, electromagnetic and hadronic calorimeters,
and a muon spectrometer incorporating three large superconducting air-core toroidal magnets.

The inner-detector system is immersed in a \qty{2}{\tesla} axial magnetic field
and provides charged-particle tracking in the range \(|\eta| < 2.5\).
The high-granularity silicon pixel detector covers the vertex region and typically provides four measurements per track,
the first hit generally being in the insertable B-layer (IBL) installed before Run~2~\cite{ATLAS-TDR-19,PIX-2018-001}.
It is followed by the SemiConductor Tracker (SCT), which usually provides eight measurements per track.
These silicon detectors are complemented by the transition radiation tracker (TRT),
which enables radially extended track reconstruction up to \(|\eta| = 2.0\).
The TRT also provides electron identification information
based on the fraction of hits (typically 30 in total) above a higher energy-deposit threshold corresponding to transition radiation.

The calorimeter system covers the pseudorapidity range \(|\eta| < 4.9\).
Within the region \(|\eta|< 3.2\), electromagnetic calorimetry is provided by barrel and
endcap high-granularity lead/liquid-argon (LAr) calorimeters,
with an additional thin LAr presampler covering \(|\eta| < 1.8\)
to correct for energy loss in material upstream of the calorimeters.
Hadronic calorimetry is provided by the steel/scintillator-tile calorimeter,
segmented into three barrel structures within \(|\eta| < 1.7\), and two copper/LAr hadronic endcap calorimeters.
The solid angle coverage is completed with forward copper/LAr and tungsten/LAr calorimeter modules
optimised for electromagnetic and hadronic energy measurements respectively.

The muon spectrometer (MS) comprises separate trigger and
high-precision tracking chambers measuring the deflection of muons in a magnetic field generated by the superconducting air-core toroidal magnets.
The field integral of the toroids ranges between \num{2.0} and \qty{6.0}{\tesla\metre}
across most of the detector.
Three layers of precision chambers, each consisting of layers of monitored drift tubes, cover the region \(|\eta| < 2.7\),
complemented by cathode-strip chambers in the forward region, where the background is highest.
The muon trigger system covers the range \(|\eta| < 2.4\) with resistive-plate chambers in the barrel, and thin-gap chambers in the endcap regions.

The luminosity is measured mainly by the LUCID--2~\cite{LUCID2} detector that records Cherenkov light produced in the quartz windows of photomultipliers located close to the beampipe.

Events are selected by the first-level trigger system implemented in custom hardware,
followed by selections made by algorithms implemented in software in the high-level trigger~\cite{TRIG-2016-01}.
The first-level trigger accepts events from the \qty{40}{\MHz} bunch crossings at a rate below \qty{100}{\kHz},
which the high-level trigger further reduces in order to record complete events to disk at about \qty{1}{\kHz}.

A software suite~\cite{SOFT-2022-02} is used in data simulation, in the reconstruction
and analysis of real and simulated data, in detector operations, and in the trigger and data acquisition
systems of the experiment.


%
\section{Data and simulated event samples}
\label{sec:mcsample}
The data sample used in this paper were collected by the ATLAS experiment during the 2015 to 2018 LHC proton--proton runs
at $\sqrt{s}=13$~\TeV, and corresponds to an integrated luminosity of $139\,\mbox{fb}^{-1}$~\cite{ATLAS-CONF-2019-021}.
Data-quality requirements are applied to ensure that all elements of the detectors were operational during data-taking~\cite{DAPR-2018-01}.
Simulated Monte Carlo (MC) samples are used to study the di-$\tau$ reconstruction and to train the di-$\tau$ identification algorithm,
while for the SF measurement, data samples are used as well.
Table~\ref{tab:MC} shows the summary of the MC generators.
For the studies of di-$\tau$ reconstruction, calibration and the training of the di-$\tau$ classifier,
simulated samples of a two-Higgs-doublet-model~\cite{Branco_2012} pseudoscalar boson production (denoted by \textit{X}) in association with a top-antitop-quark pair ($t\bar{t}X$) are used as the signal. Two representative $X$ masses ($m_{X}$) are used in the classifier training, $m_{X}=20$ and $60$~\GeV. The pseudoscalar \textit{X} was set to decay into two $\tau$-leptons, which were later set to decay hadronically. The $t\Bar{t}$ pair was set to either a semileptonic or dileptonic decay (with only electrons and muons considered).
The $t\bar{t}X$ samples were simulated at leading order (LO) with \MGNLO~\cite{Alwall:2014hca} using the \NNPDF[2.3lo] set of parton distribution functions (PDF)~\cite{Ball:2012cx},
interfaced with \PYTHIA[8.212]~\cite{Sjostrand:2014zea} to model the parton shower, hadronisation, and underlying event, with parameters set according to the A14 tune~\cite{ATL-PHYS-PUB-2014-021} and \NNPDF[2.3lo] PDF set.

Misreconstructed (fake) di-$\tau$ objects from simulated $t\bar{t}$ production (and subsequent fully hadronic decay) events are used as the background source for training the identification algorithm and SF measurement. This background is characterised by large jet multiplicities, originating from both the light- and $b$-quarks. Additionally, production of high-$\pt$ hadronically decaying $W$ bosons as part of the top-quark decays can itself result in a pair of collimated jets. The choice of the $t\bar{t}$ process as a source of fakes is further motivated by its similarity to $t\bar{t}X$.
The production of \ttbar events was modelled using the
\POWHEGBOX[v2]~\cite{Frixione:2007nw,Nason:2004rx,Frixione:2007vw,Alioli:2010xd}
generator at next-to-leading-order (NLO) in QCD with the \NNPDF[3.0nlo] PDF set~\cite{Ball:2014uwa}
and the \hdamp parameter\footnote{The
\hdamp parameter is a resummation damping factor and one of the
parameters that controls the matching of \POWHEG matrix elements to
the parton shower and thus effectively regulates the
high-\pt radiation against which the \ttbar system recoils.} set
to 1.5 $m_{\textrm{top}}$~\cite{ATL-PHYS-PUB-2016-020}.  The events were interfaced
to \PYTHIA[8.230] to model the parton shower,
hadronisation, and underlying event, with parameters set according
to the A14 tune and used the \NNPDF[2.3lo]
set of PDFs. The decays of bottom and charm hadrons
were performed by \EVTGEN[1.6.0]~\cite{Lange:2001uf}.

For the SF measurement,
the signal samples are $Z$+jets, $Z\gamma$, and $Z\gamma\gamma$ where the $Z$ boson decays into two $\tau$-leptons.
The $Z$+jets process was simulated with \SHERPA[2.2.14]~\cite{Bothmann:2019yzt}, with up to two jets at NLO and up to five jets at LO.
The $Z\gamma$ process was simulated with \SHERPA[2.2.11], with up to one jet at NLO and up to four jets at LO.
The $Z\gamma\gamma$ process was simulated with \SHERPA[2.2.10], with zero jets at NLO and up to two jets at LO.
The matrix elements were calculated with the \textsc{Comix}~\cite{Gleisberg:2008fv} and \OPENLOOPS~\cite{Buccioni:2019sur,Cascioli:2011va,Denner:2016kdg} libraries, and merged with the \SHERPA parton shower~\cite{Schumann:2007mg} following the \MEPSatNLO prescription~\cite{Hoeche:2011fd,Hoeche:2012yf,Catani:2001cc,Hoeche:2009rj} and using the set of tuned parameters developed by the \SHERPA authors.
The $Z$+jets, $Z\gamma$ and $Z\gamma\gamma$ events were simulated using the \NNPDF[3.0nnlo] PDF set~\cite{Ball:2014uwa}.
Since both the $Z$+jets and $Z\gamma$ MC samples are used, event overlap removal was performed to avoid double counting, based on the following particle-level criteria: events from $Z$+jets were discarded if they contained a photon with $\pt>140$~\GeV and if the angular distance between one of the $\tau$-leptons and the photon was greater than $0.1$.

Several SM backgrounds are used for the SF studies.
The dominant SM background is due to prompt single-photon production, which was simulated with \SHERPA[2.2.2].
In this framework, NLO matrix elements for up to two partons, and LO matrix elements for up to four partons were calculated with the \textsc{Comix} and \OPENLOOPS libraries.
They were matched with the \SHERPA parton shower using the \MEPSatNLO prescription with a dynamic merging selection~\cite{Siegert:2016bre} of \qty{20}{\GeV}.
Photons are required to be isolated according to a smooth-cone isolation criterion~\cite{Frixione:1998jh}.
The samples were simulated using the \NNPDF[3.0nnlo] PDF set, along with the dedicated set of tuned parton-shower parameters developed by the \SHERPA authors.
To estimate generator systematic uncertainties, an alternative $\gamma$+jets sample was used, with events produced at LO via \PYTHIA[8.244], using the \NNPDF[2.3lo] PDF set and the A14 tune.

The other subdominant SM backgrounds include QCD multijets with jets misreconstructed as photons or di-$\tau$ objects,
a prompt photon produced in association with a hadronically decaying $W/Z$-boson,
$t\bar{t}$, and $W$+jets production with the $W$ boson decaying leptonically.

Samples for QCD multijets production were generated with \POWHEGBOX[v2] at NLO using the dijet process, and interfaced to \PYTHIA[8.245] with the A14 tune and the \NNPDF[2.3lo] PDF set.
The \pT of the underlying Born configuration was taken as the renormalisation and factorisation scales and the \NNPDF[3.0nlo] PDF was used.
To estimate generator systematic uncertainties, an alternative multijets sample was used, with events produced at NLO with the \POWHEGBOX[v2] generator interfaced with \HERWIG[7.1]~\cite{Bellm:2015jjp}, using the \NNPDF[3.0nlo] PDF set and the default \HERWIG[7.1] tune.

The production of $W/Z(\to q\bar{q})\gamma$ was modelled by \SHERPA[2.1.1] at LO, and the parton distributions were modelled with the \CT[10] PDF set~\cite{Lai:2010vv}. The $t\bar{t}$ sample is the same as the one previously described.

The production of $W(\to\tau\nu)$+jets was simulated with the \SHERPA[2.2.1] generator using NLO matrix elements for up to two partons, and LO matrix elements for up to four partons calculated with the \textsc{Comix} and \OPENLOOPS libraries.
They were matched with the \SHERPA parton shower using the \MEPSatNLO prescription using the set of tuned parameters developed by the \SHERPA authors.
The \NNPDF[3.0nnlo] set of PDFs was used and the samples were normalised to a next-to-next-to-leading-order cross-section prediction~\cite{Anastasiou:2003ds}.

The effects of multiple proton--proton interactions in the same bunch crossing as the hard scatter and in neighbouring ones (pile-up) were included using simulated events generated with \PYTHIA[8.186] using the \NNPDF[2.3lo] PDF set and the A3 tune~\cite{ATL-PHYS-PUB-2016-017}.
Simulated events were weighted to reproduce the distribution of the average number of interactions per bunch crossing $\left<\mu\right>$ observed in data.

\begin{table}
\caption{Summary of all MC simulated samples used for the di-$\tau$ tagger development, the scale factor measurement and the estimates of the generator uncertainty assigned to the scale factor.
Information about the matrix element generator, QCD perturbative order, parton distribution function set and the parton shower is provided.
\label{tab:MC}}
\begin{center}
\begin{tabular}{lcccc}
\hline
\hline
\textbf{Process} & \textbf{Matrix element generator} & \textbf{QCD order} & \textbf{PDF}& \textbf{Parton shower}\\
\hline
\multicolumn{5}{c}{\textit{For di-$\tau$ tagger studies}}  \\
\hline
$t\bar{t}X$   & \MGNLO  & NLO & \NNPDF[2.3nnlo] & \PYTHIA[8.212] \\
$t\Bar{t}$ & \POWHEGBOX[v2]      & NLO & \NNPDF[3.0nlo] & \PYTHIA[8.230] \\
\hline
\multicolumn{5}{c}{\textit{For scale factor measurement}} \\
\hline
$Z$+jets                &\SHERPA[2.2.14] & NLO & \NNPDF[3.0nnlo] & \SHERPA \\%
$Z\gamma$              &\SHERPA[2.2.11] & NLO & \NNPDF[3.0nnlo] & \SHERPA \\%
$Z\gamma\gamma$  &\SHERPA[2.2.10] & NLO & \NNPDF[3.0nnlo] & \SHERPA \\%
$\gamma$+jets           &\SHERPA[2.2.2]    & NLO & \NNPDF[3.0nnlo] & \SHERPA \\%
$W/Z(\to q\bar{q})\gamma$            &\SHERPA[2.1.1]  & LO  & CT10 & \SHERPA \\%
Multijet          &\POWHEGBOX[v2]  & NLO & \NNPDF[3.0nlo] & \PYTHIA[8.245] \\
$t\Bar{t}$              &\POWHEGBOX[v2]  & NLO & \NNPDF[3.0nlo] & \PYTHIA[8.230] \\
$W(\to\tau\nu)$+jets &\SHERPA[2.2.1]  & NLO & \NNPDF[3.0nnlo] & \SHERPA \\%
\hline
\multicolumn{5}{c}{\textit{For generator uncertainty estimates}} \\
\hline
$\gamma$+jets           &\PYTHIA[8.244]+ \EVTGEN[1.7.0] & LO & \NNPDF[2.3lo] & \PYTHIA[8.244] \\
Multijet          &\POWHEGBOX[v2]  & NLO & \NNPDF[3.0nlo] & \HERWIG[7] \\
\hline
\hline
\end{tabular}
\end{center}
\vspace{10pt} %
\end{table}


%
\section{Event reconstruction}
\label{sec:reconstruction}
The following procedures are used to reconstruct photons, electrons, muons, jets, large-radius (large-$R$) jets, and the missing transverse momentum.

Photons and electrons are reconstructed from clusters of energy deposits in the EM calorimeter,
together with tracks reconstructed in the inner tracking detector~\cite{EGAM-2018-01,EGAM-2021-01,EGAM-2021-02}.
Photon candidates are required to have $\pt>150\,\GeV$ and $|\eta|<2.37$.
The identification (ID) of photons is performed by requiring the photon to satisfy a set of identification criteria~\cite{EGAM-2018-01}  %
based on shower shapes measured in the first two longitudinal layers of the electromagnetic calorimeter,
where the first layer has high granularity and provides large discrimination between prompt photons and photons from decays of hadrons inside jets,
and the leakage into the hadronic calorimeter.
Two isolation (ISO) working points (WP) of photons, \textit{Tight} and \textit{Loose}~\cite{EGAM-2018-01}, are defined
based on the amount of transverse energy deposited in clusters of calorimeter cells within a cone of radius $R=0.4$ and $R=0.2$, respectively, around the photon, excluding the photon cluster itself,
and the track isolation within a cone of radius $R=0.2$ around the photon.

Electron candidates are identified using the likelihood identification criteria described in Ref.~\cite{EGAM-2018-01}.
The \textit{VeryLoose} identification criteria are applied to electrons.
Candidates are required to have $\pt>10$~\GeV and $|\eta| < 2.47$.

Muon candidates are reconstructed from tracks in the muon spectrometer that are matched to a corresponding track in the inner tracking detector~\cite{MUON-2018-03}.
Candidates are required to have $\pt>10$~\GeV, $|\eta| < 2.7$, and satisfy the \textit{Loose} identification criteria. %

Jets are reconstructed from particle-flow objects~\cite{PERF-2015-09} using the anti-$\kt$ jet algorithm with a radius parameter of $R=0.4$.
The jets are calibrated following the procedure described in Ref.~\cite{JETM-2018-05} and are required to have $\pt > 30$~\GeV and $|\eta| < 2.5$.
Pile-up jets with $\pt <$ 120~\GeV are rejected using the \textit{Medium} working point of the jet vertex tagger~\cite{PERF-2014-03}.

Large-$R$ jets are reconstructed using the anti-$\kt$ jet algorithm with a radius parameter of $R=1.0$ from particle-flow objects.
Candidates are required to have $\pt>50$~\GeV and $|\eta|<2.5$, to ensure good overlap with the tracking volume of the ATLAS detector and reduce pile-up jet contamination.
The missing transverse momentum (with magnitude \MET) is reconstructed as the negative vector sum of the transverse momenta of all the reconstructed and calibrated objects in the event, including a soft term that accounts for all tracks associated with the primary vertex but not matched to any reconstructed object~\cite{JETM-2020-03}.

A standard overlap-removal procedure is applied to resolve ambiguities where multiple physical objects are reconstructed from the same detector signature.
Additionally, in the SF measurement, an overlap removal between photon and di-$\tau$ objects is performed, prioritizing the photons within $\Delta R=1.0$ of the di-$\tau$ candidate. In a more general context, the di-$\tau$ object's priority in the object overlap removal hierarchy would place it below light leptons and photons, and above jets.

%
%
%

%
%
%
%
%
%
%
%
%
%
%
%
%
%
%
%
%
%
%

%
%
%
%
%
%
%
%
%
%
%


%
\section{Di-$\tau$ reconstruction, energy-scale calibration, and identification}
\label{sec:ditaureconstruction}
\begin{figure}
\centering
\subfigure[]{\centering\includegraphics[height=0.5\textwidth]{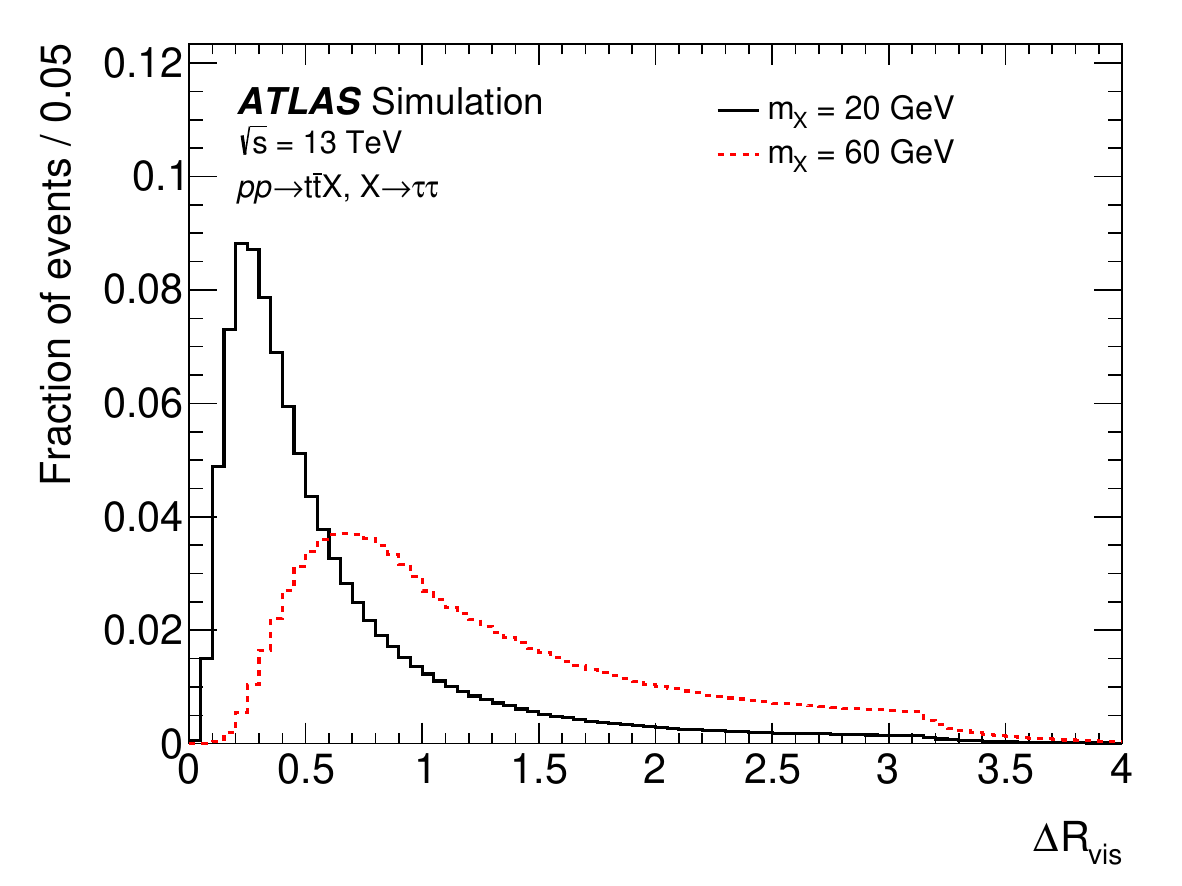}
\label{fig:drtautau}}
\subfigure[]{\centering\includegraphics[width=0.45\textwidth]{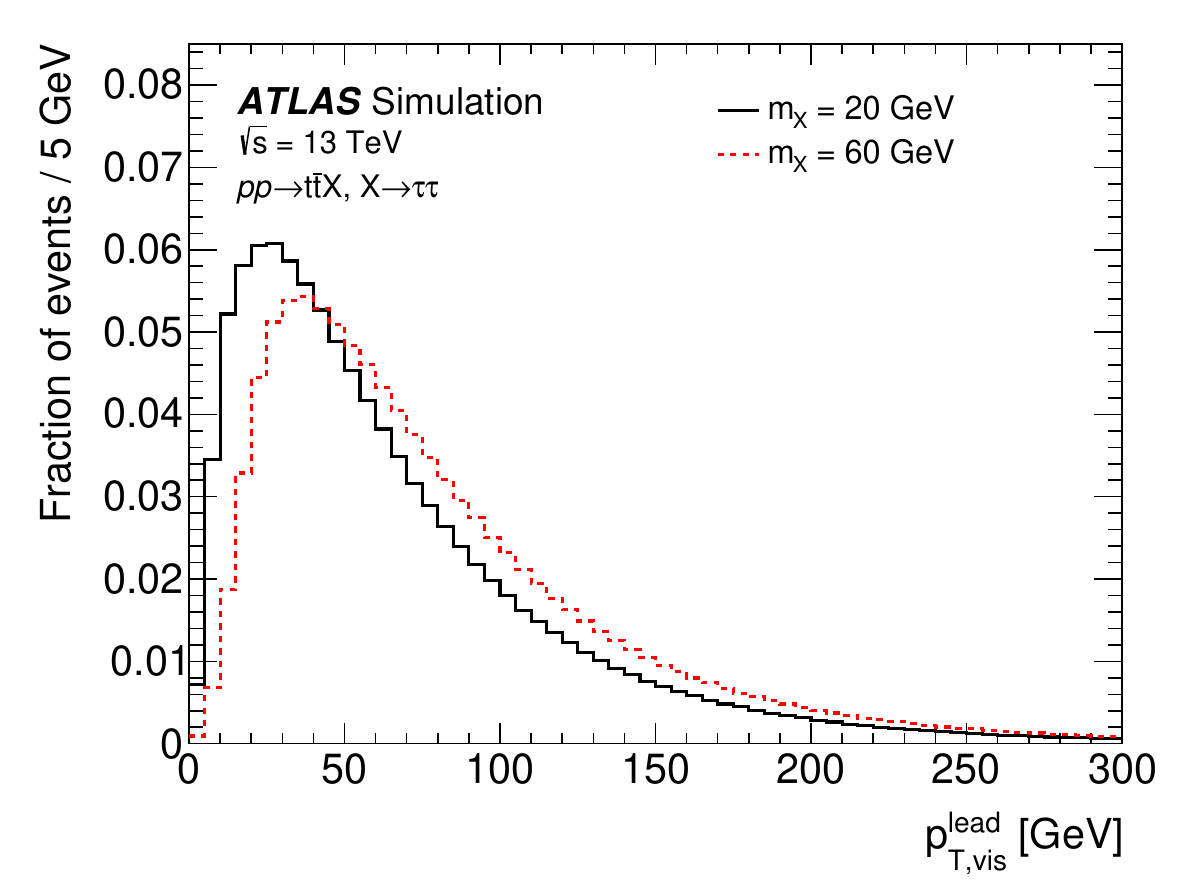}
\label{fig:tauptlead}}
\subfigure[]{\centering\includegraphics[width=0.45\textwidth]{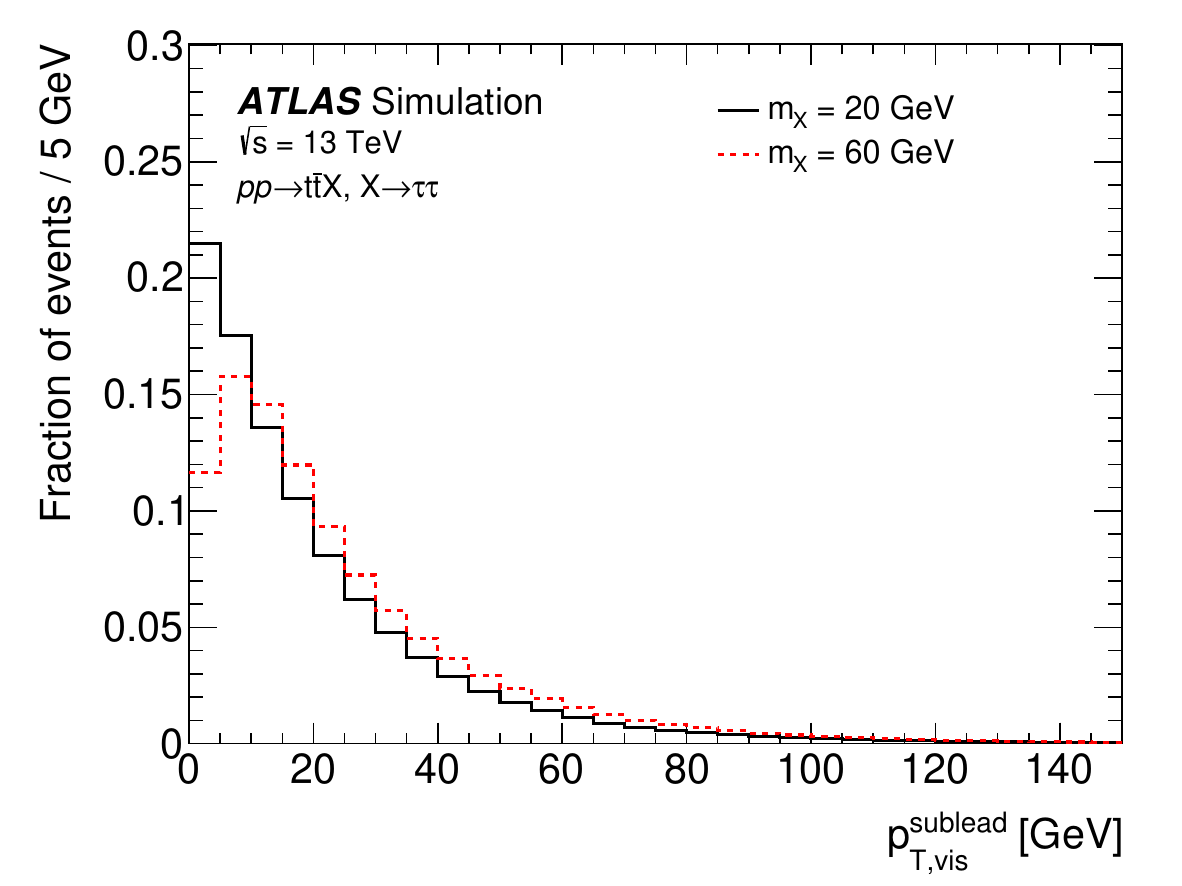}
\label{fig:tauptsubl}}
\caption{Distributions of (a) visible angular distance $\Delta R_{\text{vis}}$, (b) leading \tauhadvis \pt and (c) subleading \tauhadvis \pt in
$X\to \tauhad \tauhad$ decays from $t\Bar{t}X$ events,
normalised to unit area, for two different $m_X$ values.}
\end{figure}

One of the major challenges facing light resonance searches~\cite{HDBS-2018-47,CMS-HIG-18-024,ATLAS:2021hbr} is the tagging of the resonance decay products. Due to the low mass of the \textit{X} resonance, a significant fraction of the \textit{X} resonances will be produced with transverse momenta sizeable enough to result in decay products that are highly collimated. This is demonstrated in Figure~\ref{fig:drtautau}, showing the distributions of $\Delta R_{\text{vis}}$, the particle-level visible angular distance between the two \tauhadvis, in simulated $t\bar{t}X$ ($X\rightarrow\tauhad\tauhad$) events for $m_X$ values of 20~\GeV and 60~\GeV. In the following, the notations of \textit{leading} and \textit{subleading} refer to their ordering in \pt.

\subsection{Reconstruction}
\label{sec:reco}
The tagging of a nearby \tauhadvis pair relies on the reconstruction of a large-$R$ jet (seed jet) and its substructure. The reconstruction algorithm was initially developed for boosted $h\rightarrow\tauhad\tauhad$ decays, in the context of a search for resonant di-Higgs boson production in the $b\Bar{b}\tau^+\tau^-$ channel~\cite{HDBS-2019-22}. In that case, the \tauhadvis were expected to be produced with large individual transverse momenta, and the reconstruction was only performed for seed jets with $\pt>300$~\GeV. However, when a \textit{light} resonance decays into a boosted di-$\tau$, the \tauhadvis $\pt$ spectrum is rather soft, as demonstrated in Figures~\ref{fig:tauptlead} and~\ref{fig:tauptsubl}. In accordance, the seed jet $\pt$ threshold is reduced down to 50~\GeV, and the di-$\tau$ objects reconstructed from those low-$\pt$ seeds are later studied.

\begin{figure}
\centering
\includegraphics[width=0.45\linewidth]{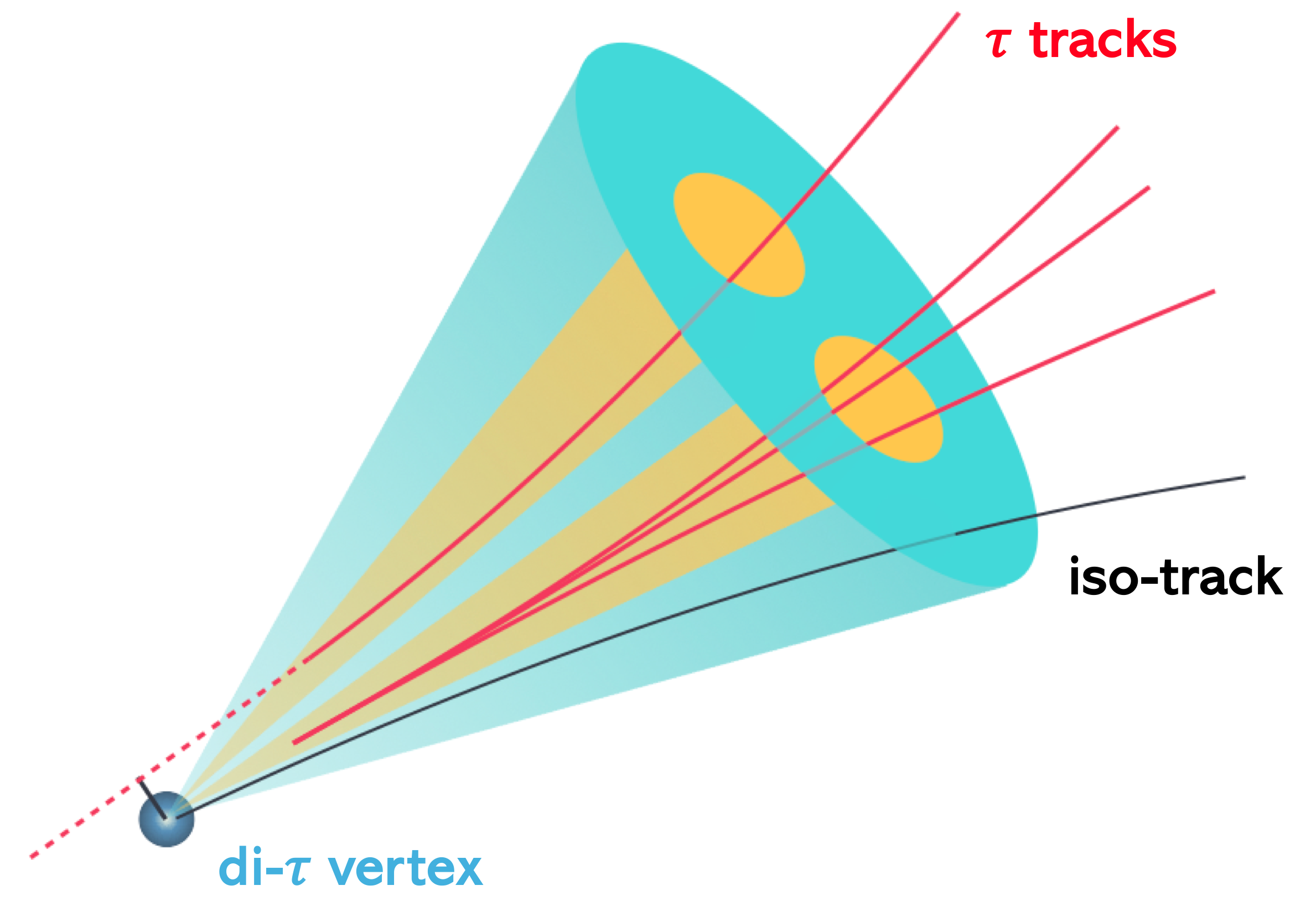}
\caption{Schematic illustration of the reconstructed di-$\tau$ object topology~\cite{HDBS-2019-22} for one seed jet with $R=1.0$ and two subjets with $R=0.2$.
\label{fig:ditautopo}}
\end{figure}

During Run 1 and early Run 2 of the LHC, \tauhadvis candidates reconstructed from $R=0.4$ anti-$\kt$ jets in the ATLAS experiment utilised the definitions of an inner cone of radius $R=0.2$ (core cone) and its surrounding annulus $0.2<R<0.4$ (isolation annulus)~\cite{PERF-2013-06}.
Once a seed jet is reconstructed, its constituents are grouped into subjets (with $\pt>10$~\GeV) using the anti-$\kt$ algorithm with $R=0.2$. Thus, a consistent definition with the single \tauhadvis core cone, where the signature from its decay products is expected, is maintained.
The seed jet area not included within the radius of any subjet (containing tracks and energy deposits not associated with any subjet) is considered as the isolation region, analogous to the isolation annulus of a \tauhadvis candidate.
A core region is also defined for each subjet as the cone of radius $R=0.1$ around its axis. An illustration of the reconstructed di-$\tau$ object topology is given in Figure \ref{fig:ditautopo}.
A hadronic di-$\tau$ candidate is required to have at least two subjets, each with at least one associated track. A di-$\tau$ vertex calculation is performed
to find the most likely di-$\tau$ production vertex, and subjet kinematics are calculated relative to this vertex. The track selection and track-to-vertex matching criteria are identical to those used in Ref.~\cite{ATL-PHYS-PUB-2015-045}.
Impact parameter requirements used in associating tracks to subjets are calculated relative to the di-$\tau$ vertex, while for isolation region tracks the vertex with the highest $\sum(\pt^{\text{trk}})^2$ is used. In the following, it is assumed that the two leading subjets hold the \tauhadvis signatures and the sum of their four-momentum defines that of the \tauhad pair. This assumption is valid in approximately 90\% of cases for the considered $t\bar{t}X$ samples where the particle-level \tauhadvis are both captured by any two unique subjets of a reconstructed di-$\tau$ object.

A truth-matched di-$\tau$ is defined as a reconstructed di-$\tau$ object in which the leading and subleading subjets are each geometrically-matched to particle-level \tauhadvis within $\Delta R=0.2$. The di-$\tau$ reconstruction efficiency is then defined as the fraction of events containing a truth-matched di-$\tau$ object out of all events satisfying a baseline selection applied at particle-level.

\begin{figure}
\centering
\subfigure[]{\includegraphics[width=0.45\textwidth]{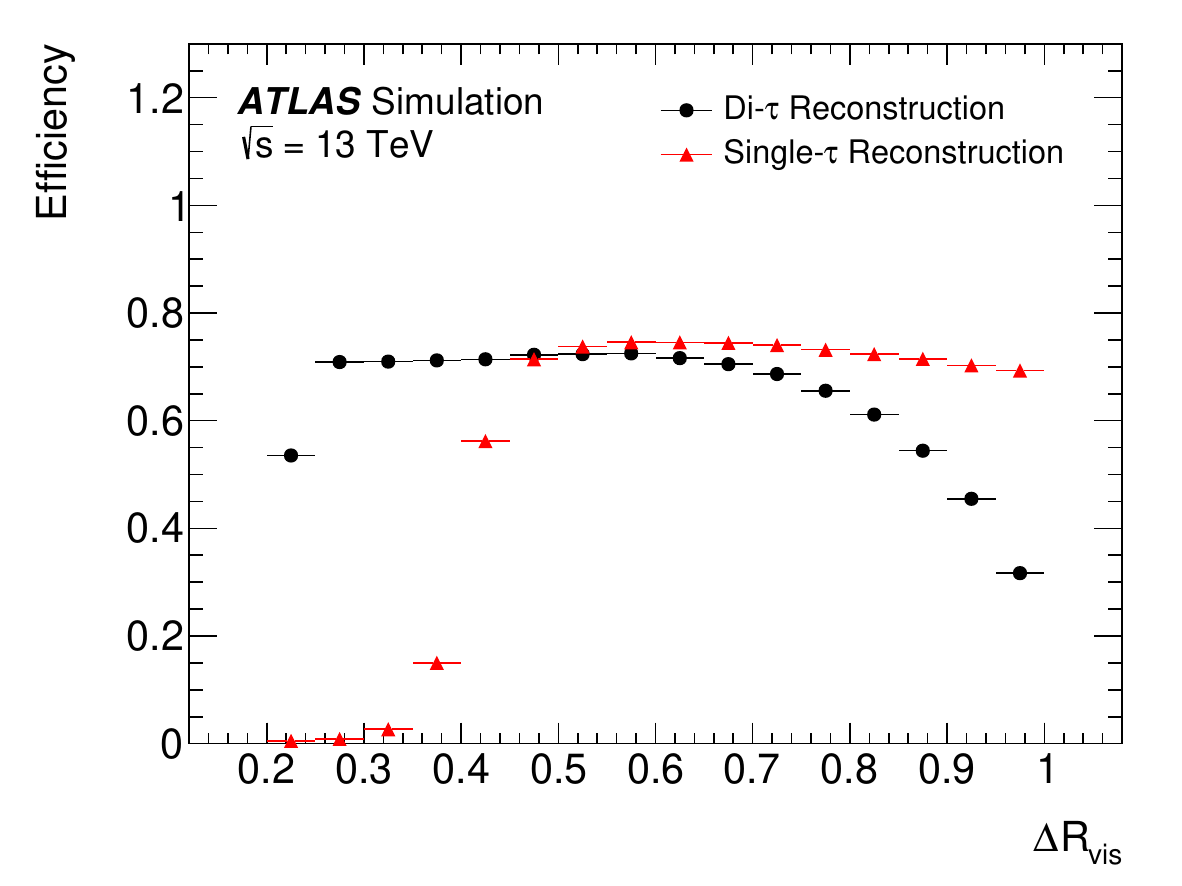}}
\subfigure[]{\includegraphics[width=0.45\textwidth]{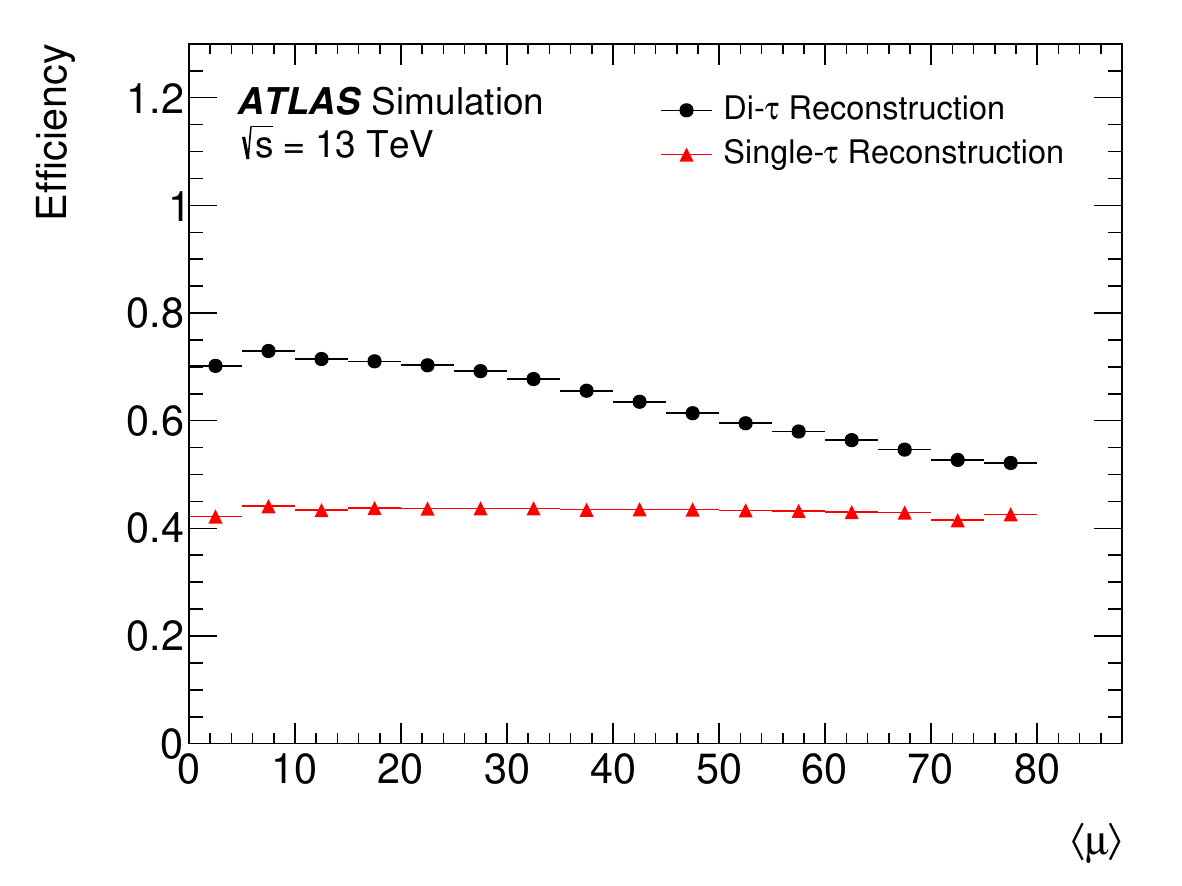}}
\caption{Di-$\tau$ reconstruction efficiency (dots) and two resolved~\tauhadvis reconstruction efficiency (triangles) as a function of (a) the visible angular separation $\Delta R_{\text{vis}}$ between the two particle-level \tauhadvis and (b) the average number of interactions~$\left<\mu\right>$, measured in simulated $t\Bar{t}X$ events. The $t\Bar{t}X$ sample includes events with $m_X$ = 20~\GeV and $m_X$ = 60~\GeV in equal proportions. The plateau value in (b) for the two resolved~\tauhadvis case depends strongly on the fraction of events with angular distance smaller than 0.4, which limits the reconstruction efficiency achievable using the resolved~\tauhadvis method, and hence represents an averaged efficiency between the two $m_X$ values. The error bars account for the statistical uncertainty in simulation.
\label{fig:recoeff}}
\end{figure}

To satisfy the baseline selection, an event is required to have a single particle-level hadronically decaying $\tau^+\tau^-$ pair originating from the \textit{X} resonance (discarding  \tauhad leptons from heavy-flavour hadron decays), with an angular distance $0.2\leq\Delta R_{\text{vis}}\leq1.0$ between the two \tauhadvis, each of which is required to have $p_{\text{T,vis}}\geq10$~\GeV. These criteria are selected to reflect the phase-space defined by the previously described subjet reconstruction step.
The reconstruction efficiencies for the boosted di-$\tau$ tagger and for two standard (resolved) \tauhadvis are calculated in $t\bar{t}X$ events, and are given in Figure \ref{fig:recoeff}.
It is evident from the figure that the di-$\tau$ reconstruction is indeed most efficient in the highly collimated $\Delta R_{\text{vis}}\leq0.45$ regime, where resolved~\tauhadvis reconstruction fails, and an overall larger fraction of events may be successfully reconstructed. However, its performance is clearly less resilient to increasing pile-up conditions, degrading by about 20\% over the examined range. This is mainly due to the soft spectrum of the subleading \tauhadvis making it increasingly probable -- as conditions become more severe -- that pile-up contributions cause significant shifts in reconstructed seed jet and subjet axes, such that both the particle-level \tauhadvis are not successfully captured inside the two leading subjets of a single seed jet. Similarly, as the $p_{\text{T,vis}}$ distribution becomes softer and the angular separation between the two \tauhadvis increases, a single $R=1.0$ seed jet and its subjets are less likely to capture both the \tauhadvis and the reconstruction efficiency correspondingly declines.
Additionally, as the individual \tauhadvis charged-hadron multiplicities increase, the di-$\tau$ reconstruction efficiency increases by approximately 10\%, while having an alleviated dependence on pile-up conditions.

\subsection{Energy scale calibration}
\begin{figure}
\centering
\subfigure[]{\includegraphics[width=0.45\linewidth]{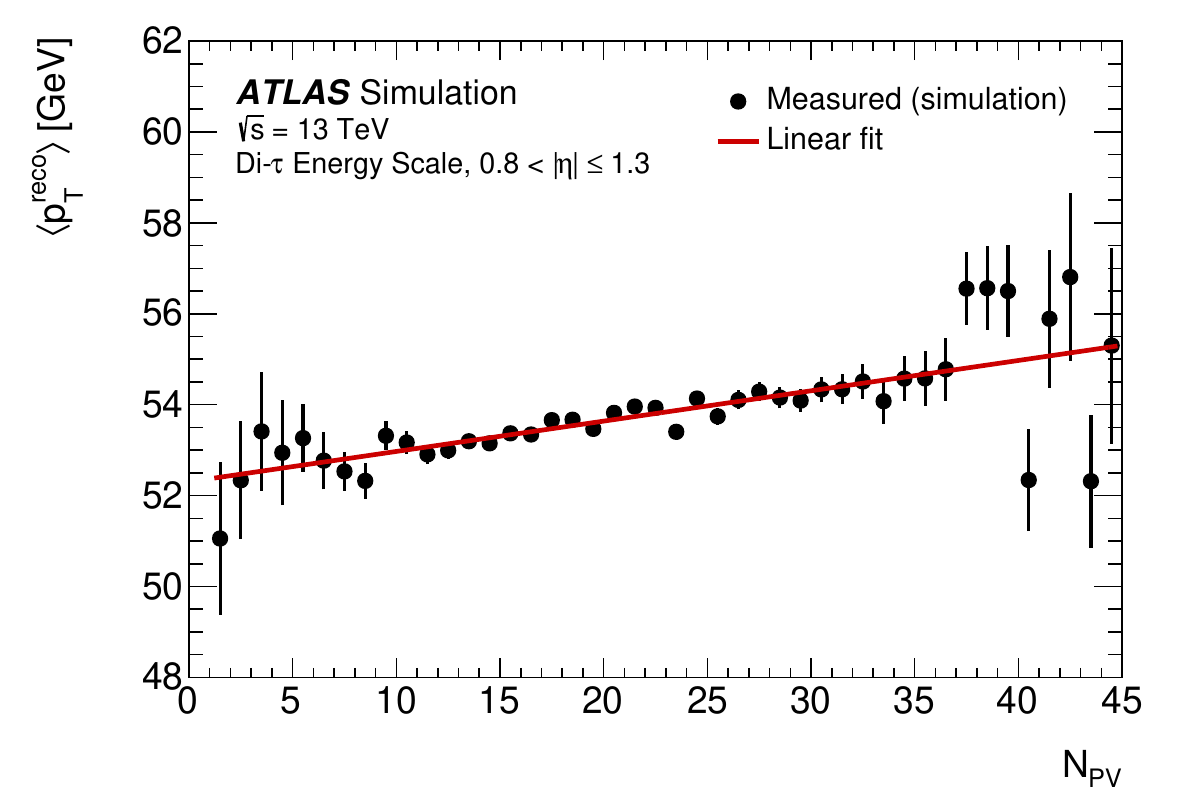}\label{fig:puprof}}
\subfigure[]{\includegraphics[width=0.45\linewidth]{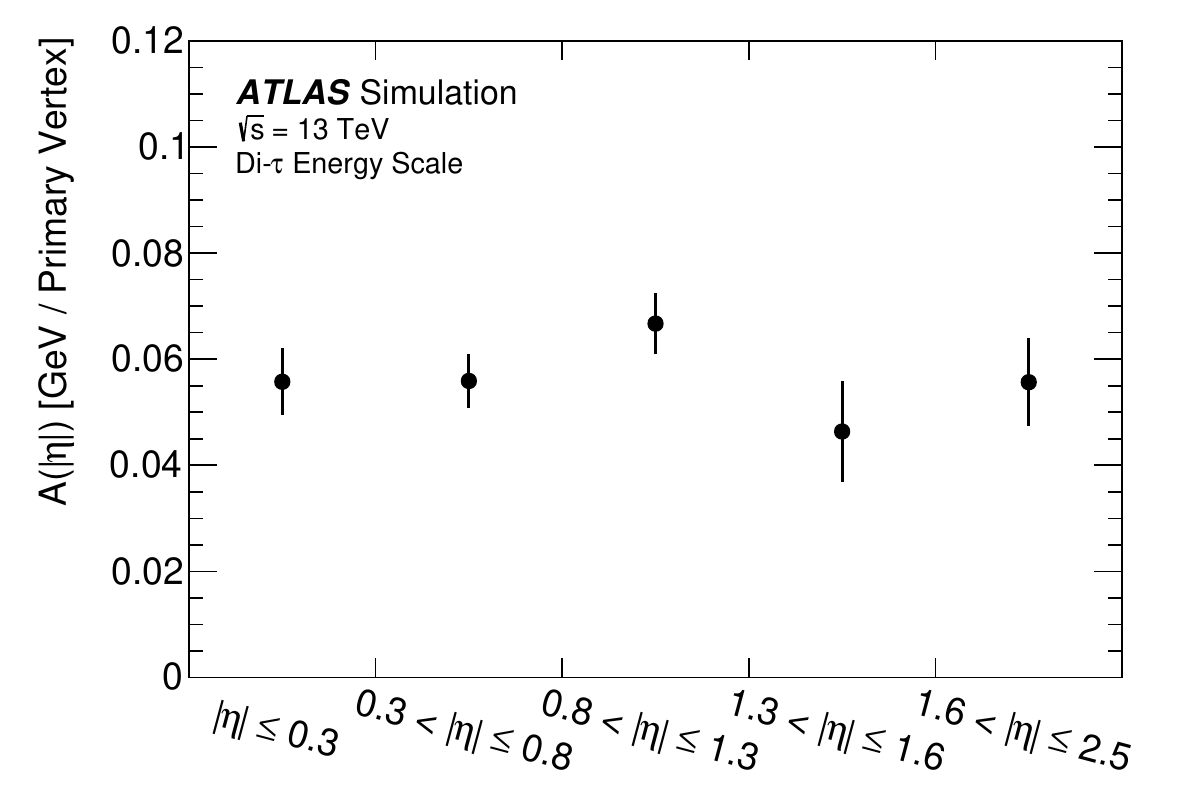}\label{fig:PUcoeffs}}
\caption{(a) Mean subjet $\pt^{\text{reco}}$ as a function of the number of reconstructed primary vertices for $0.8<|\eta_{\text{reco}}|\leq1.3$, where the line is the linear fit from which correction coefficients are derived. (b) Measured linear pileup-correction coefficients $A(|\eta_{\text{reco}}|)$. These are obtained for \tauhadvis originating from $X$ decays in $t\Bar{t}X$ events, using the two generated $m_X$ values. The error bars in (a) account for the statistical uncertainty in simulation, while the error bars in (b) represent the uncertainty on the linear fit slope parameter.}
\end{figure}

\begin{figure}
\centering
\subfigure[]{\includegraphics[height=0.35\textwidth]{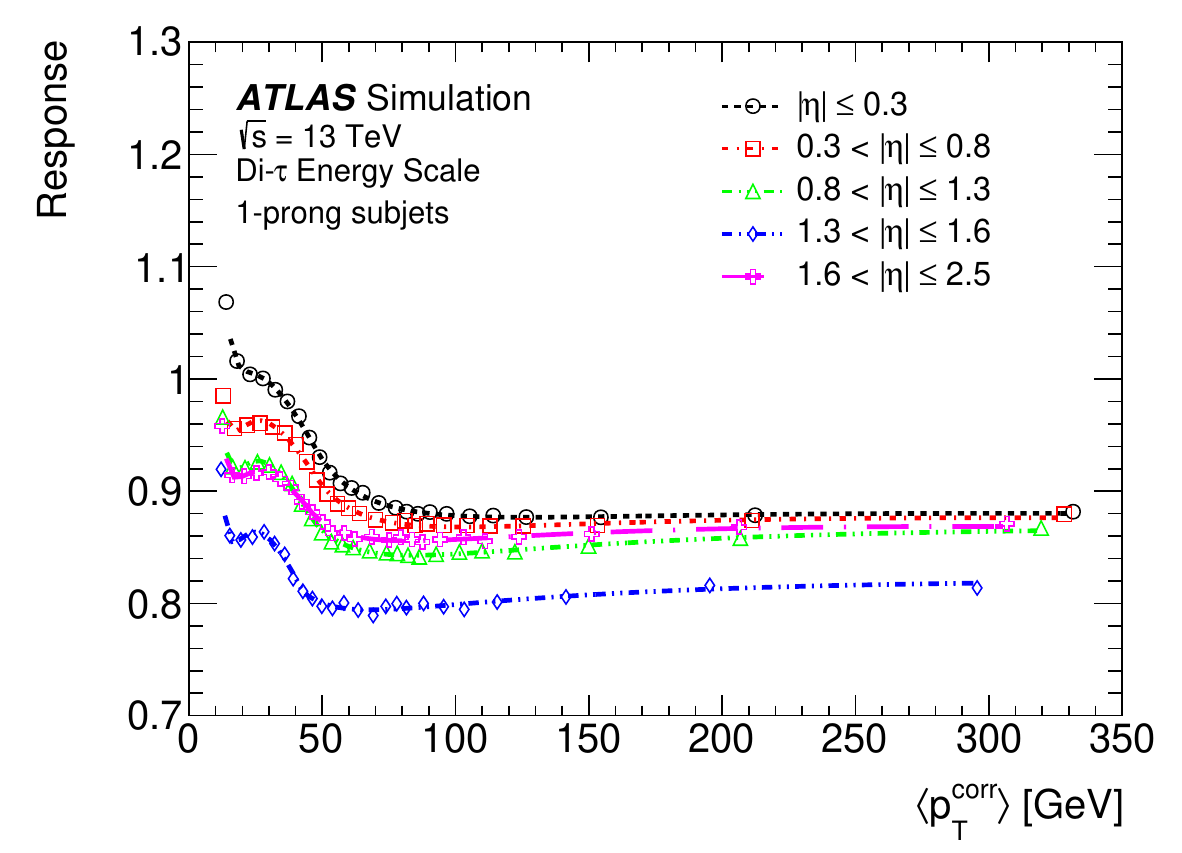}}
\subfigure[]{\includegraphics[height=0.35\textwidth]{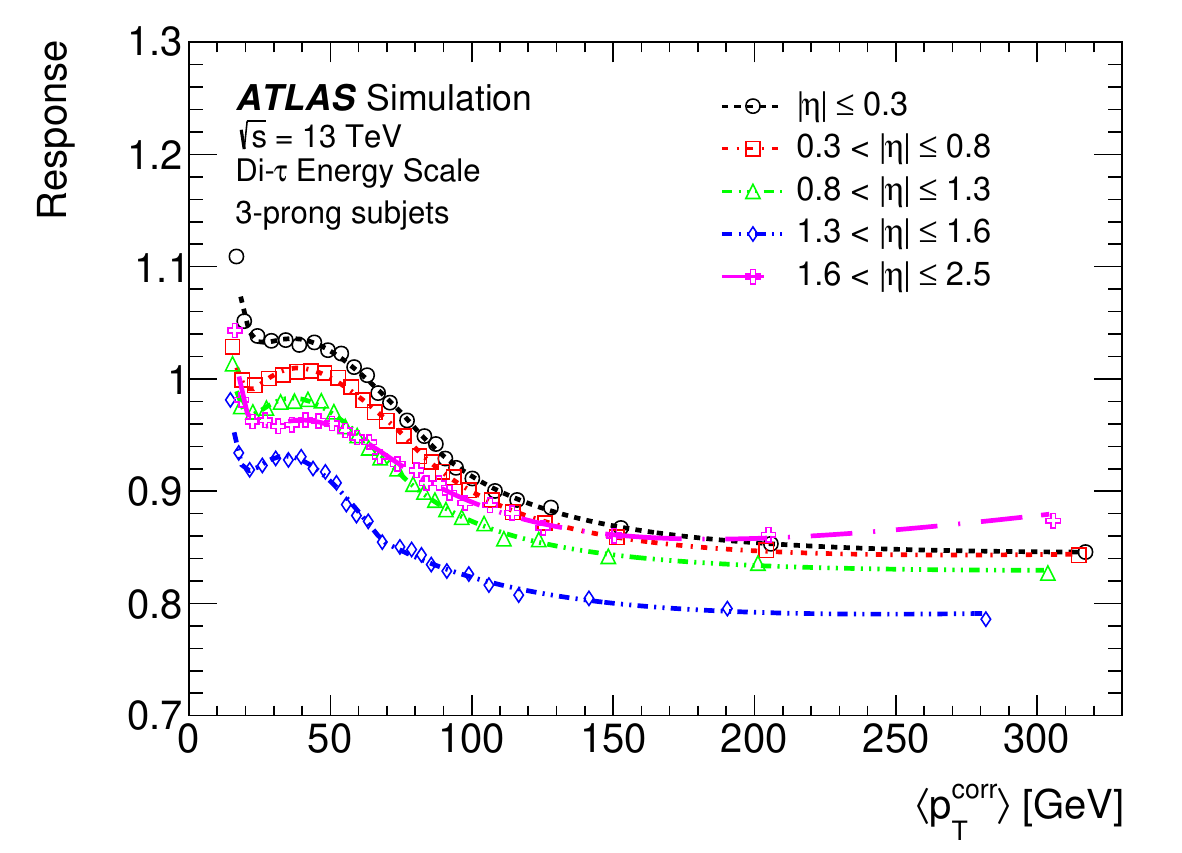}}
\caption{Detector response function for (a) 1-prong and (b) 3-prong subjets as a function of $\left<\pt^{\text{corr}}\right>$ in various $|\eta_{\text{reco}}|$ regions, for \tauhadvis originating from $X$ decays in $t\Bar{t}X$ events, generated with $m_X$ values of 20~\GeV and 60~\GeV.
\label{fig:respfunc}}
\end{figure}

Truth-matched di-$\tau$ objects from the two generated $m_X$ points, with either one or three charged tracks associated with each of the two leading subjets, are later used to compare reconstructed and particle-level $\pt$ values; from this comparison, corrections are derived to calibrate the reconstructed momentum to the particle-level \tauhadvis scale. The calibration is conducted individually for each subjet, binned in reconstructed $|\eta|$ ($|\eta_{\text{reco}}|$) and associated track multiplicity ($N_{\text{prong}}$), in a two-step procedure similar to the one described in Ref.~\cite{ATL-PHYS-PUB-2015-045}. In the first step, the contribution to the subjet momentum due to pile-up interactions is estimated and subtracted.
In the second step, a detector response correction is applied,
aiming to provide the best estimate of the true \tauhadvis momentum.
As demonstrated in Figure~\ref{fig:puprof}, the subjet \pt is found to increase linearly with the number of reconstructed primary vertices ($N_{\text{PV}}$) in all $|\eta_{\text{reco}}|$ regions, with each vertex adding around 60~\MeV to the measured $\pt$. The pile-up corrected momentum is thus given by: $\pt^{\text{corr}}=\pt^{\text{reco}}-A(|\eta_{\text{reco}}|)\times N_{\text{PV}}$. The pile-up-correction coefficients $A(|\eta_{\text{reco}}|)$ are summarised in Figure \ref{fig:PUcoeffs}. In the second step, a detector response function is derived from the ratio of corrected and generated visible momentum in each $|\eta_{\text{reco}}|$ and $N_{\text{prong}}$ region. This function is denoted $R(\pt^{\text{corr}},|\eta_{\text{reco}}|,N_{\text{prong}})$, and is used to derive the calibrated momentum as:
$\pt^{\text{calib}}=\frac{\pt^{\text{corr}}}{R(\pt^{\text{corr}},|\eta_{\text{reco}}|,N_{\text{prong}})}$. Figure~\ref{fig:respfunc} shows the detector response functions. For $\pt$ values greater or lower than the measured points, the response function is set to a constant value (its value at the measured limit). Response function values are mostly below unity due to the underlying EM scale calibration of constituent particle flow objects. The response generally displays lower values as $|\eta_{\text{reco}}|$ increases, except the $1.3<|\eta_{\text{reco}}|\leq1.6$ region, where the transition between barrel and end-cap calorimeters occurs and even lower response values are observed.
The subjet \pt threshold implemented in the reconstruction induces a bias in the response distribution of very low-\pt \tauhadvis, favouring candidates reconstructed with a high response and leading to a larger average response at $\left<\pt^{\text{corr}}\right>\lesssim 20$~\GeV.
As larger associated track multiplicities provide better momentum estimations at low $\pt^{\text{corr}}$, the response values for 3-prong subjets are closer to unity than for 1-prong subjets across a larger $\left<\pt^{\text{corr}}\right>$ range.


\subsection{Identification}
The di-$\tau$ reconstruction method does not provide background separation power. Limited rejection can be obtained from the introduction of selection criteria, for example on the number of subjets and their associated track multiplicities and charges. To further discriminate genuine boosted \tauhadvis pairs from misreconstructed di-$\tau$ candidates (originating from jets), a dedicated identification step is introduced. Identification variables are calculated for each di-$\tau$ candidate, using tracking and calorimeter information from the subjets and the isolation region. These variables are later used as inputs to train a boosted decision tree (BDT)~\cite{Breiman:2017lcz,Freund:1996:ENB:3091696.3091715,Freund:1997xna} classifier.
The signal for the classifier consists of truth-matched di-$\tau$ objects from the previously mentioned $t\bar{t}X$ events using the two generated $m_X$ points, chosen as the upper and lower limits of the $m_X$ range. When combined, these mass points provide a di-$\tau$ $p_{\text{T,vis}}$ phase space coverage between approximately 50--250~\GeV at the most relevant $0.2\leq\Delta R_{\text{vis}}\leq0.6$ range, where the reconstruction efficiency is greatest. The background is composed of fake di-$\tau$ objects originating from jets in all-hadronic $t\Bar{t}$ decay events. Di-$\tau$ objects entering the training are required to have either one or three charged-particle tracks associated with each of the two leading subjets.

The particular set of BDT input variables was chosen by reducing a larger set of calculated variables in steps, with consideration taken to include variables that contain information from all regions of a reconstructed di-$\tau$: the core cone of both of the subjets, the full area of both of the subjets, the isolation region and the entire seed jet. The bulk of signal di-$\tau$ objects have \pt values in the range of approximately 70 to 150~\GeV, while the background \pt spectrum is softer, ranging from approximately 20 to 80~\GeV. To mitigate the dependence of the BDT output score on the transverse momentum, input events are reweighted such that the resulting di-$\tau$ $\pt$ spectrum (separately for signal and background) is flattened up to 250~\GeV, and exponentially decreasing beyond.

The 16 variables used as input to the classifier are:\footnote{The notation (sub)lead refers to the (sub)leading $\pt$ subjet within the seed jet.}
\begin{itemize}
\item $n_{\text{isotracks}}$: the number of tracks associated with the isolation region.
\item Subjet \pt fraction $f_{\text{subjet}}^{\text{(sub)lead}}$: the ratio between the transverse momenta of the subjet and the seed jet,
$$f_{\text{subjet}}^{\text{(sub)lead}} \equiv \frac{\pt^{\text{(sub)lead}}}{\pt^{\text{seed}}}.$$
\item $R_{\text{isotrack}}$: the $\pt$-weighted sum of track distances from the subjet axis, for isolation-region tracks  inside a cone of $\Delta R<0.4$ around the subjets,
$$R_{\text{isotrack}} \equiv \frac{\sum_{\text{(sub)lead}}\sum_{i}^{\Delta R_i<0.4}p_{\text{T},i}^{\text{isotrk}}\Delta R_i}{\sum_{\text{(sub)lead}}\sum_{i}^{\Delta R_i<0.4}p_{\text{T},i}^{\text{isotrk}}}.$$
This definition means the variable considers only tracks not associated with any subjet, that are within an isolation annulus similar to that of a single reconstructed \tauhadvis. A value of zero is assigned if no tracks are found.
\item $R_{\text{max}}^{\text{(sub)lead}}$: the maximal $\Delta R$ of an associated track to the subjet axis.
\item Weighted core track distance $R_{\text{core}}^{\text{(sub)lead}}$: defined for a given subjet, this is the $\pt$-weighted sum of track distances from the subjet axis, for tracks found inside the core cone of the subjet,
$$R_{\text{core}}^{\text{(sub)lead}} \equiv \frac{\sum_{i}^{\Delta R_i<0.1}p_{\text{T},i}^{\text{trk}}\Delta R_i}{\sum_{i}^{\Delta R_i<0.1}p_{\text{T},i}^{\text{trk}}}.$$
A value of zero is assigned if no tracks are found inside the core cone.
\item $R_{\text{track}}$: $\pt$-weighted sum of track distances from the subjet axis, for \tauhadvis tracks  inside a cone of $\Delta R<0.2$ around the subjets,
$$R_{\text{track}} \equiv \frac{\sum_{\text{(sub)lead}}\sum_{i}^{\Delta R_i<0.2}p_{\text{T},i}^{\text{trk}}\Delta R_i}{\sum_{\text{(sub)lead}}\sum_{i}^{\Delta R_i<0.2}p_{\text{T},i}^{\text{trk}}}.$$
\item $f_{\text{track}}^{\text{(sub)lead}}$: the ratio between the highest-$\pt$ track inside a subjet, and the respective subjet $\pt$.
\item $\log(m_{\text{tracks}}^{\text{(sub)lead}})$: logarithm of the invariant mass calculated from the four-momenta of tracks associated with the given subjet.
\item $\log(|d_{0,\text{lead-track}}^{\text{(sub)lead}}|)$: logarithm of the closest distance in the transverse plane between the primary vertex and the leading track associated with the appropriate subjet.
\item $\Delta R(\text{lead,\ sublead})$: angular separation between the two leading subjets.
\end{itemize}

\begin{figure}
\centering
\subfigure[]{\centering\includegraphics[width=0.45\textwidth]{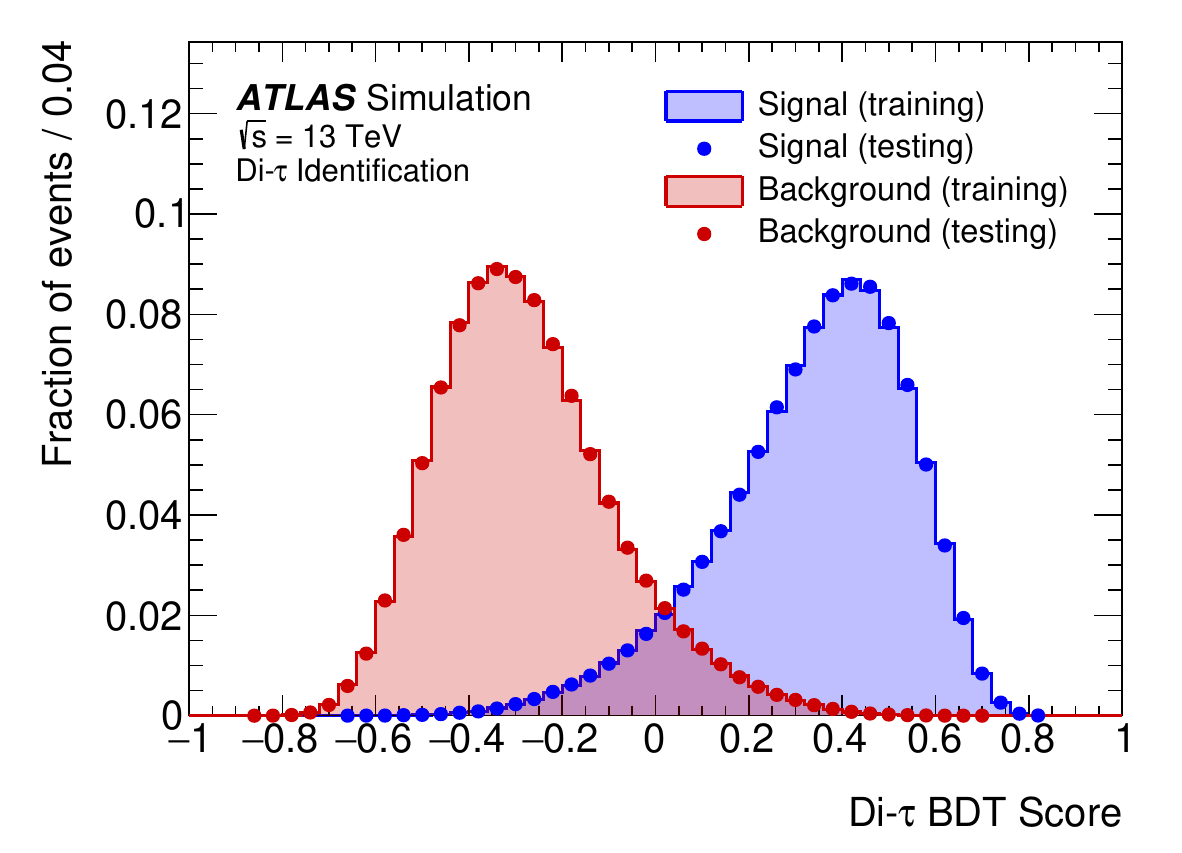}\label{fig:overtrain}}
\subfigure[]{\centering\includegraphics[width=0.45\textwidth]{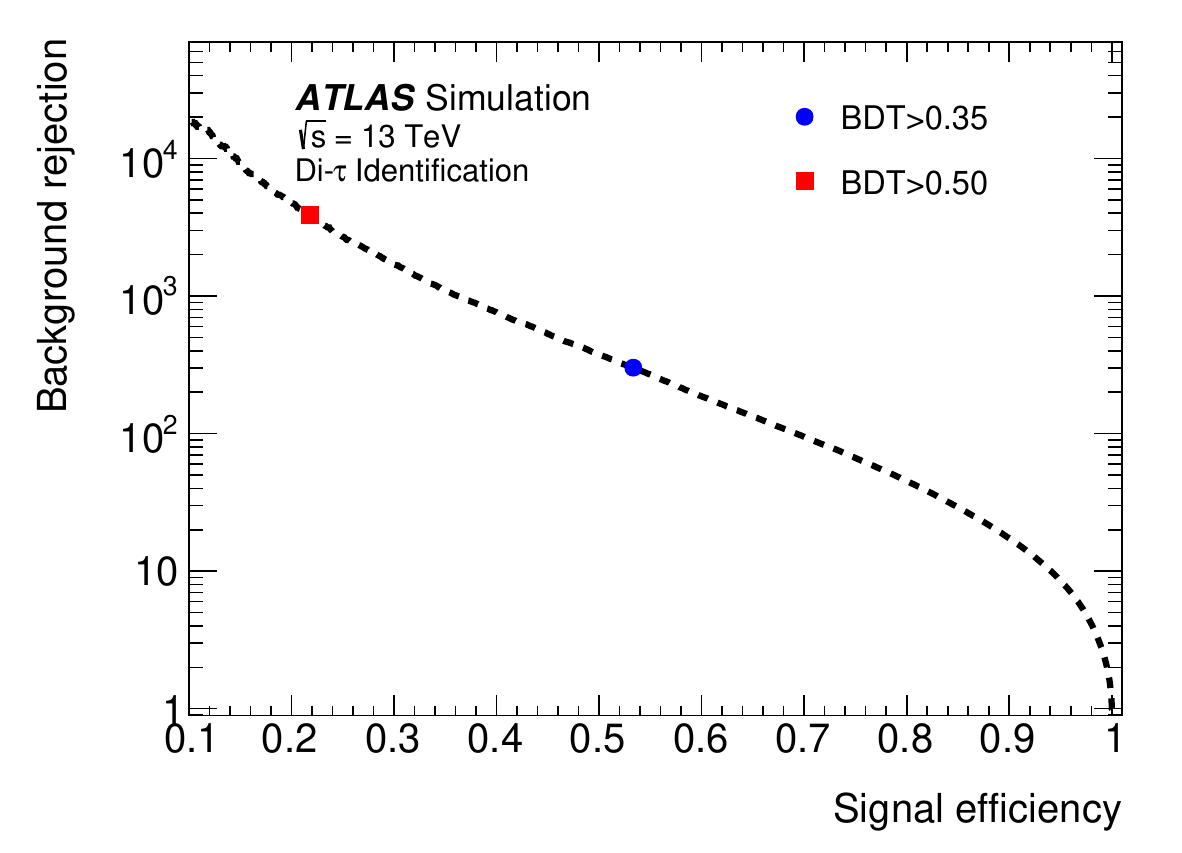}\label{fig:ROC}}
\caption{Results of the classifier training, showing the (a) BDT score distributions for signal and background events and (b) background rejection factor versus signal identification efficiency, corresponding to the trained BDT score distributions. The two markers represent the \textit{Medium} and \textit{Tight} WPs. Signal (real) di-$\tau$ objects originate from $X$ decays in $t\Bar{t}X$ events, generated with $m_X$ values of 20~\GeV and 60~\GeV in equal proportions, while background (fake) di-$\tau$ objects originate from jets in fully-hadronic $t\Bar{t}$ events.}
\end{figure}
The resulting BDT distributions for training and testing events (for both signal and background) are presented in Figure \ref{fig:overtrain}, showing two well-separated peaks for the signal and background with no evidence of classifier overtraining. The resulting separation power is better illustrated through the Receiver Operating Characteristic (ROC) curve, which is defined here as the inverse background efficiency (background rejection) as a function of signal efficiency. The ROC curve for the trained BDT is given in Figure \ref{fig:ROC}, corresponding to an area under curve of 0.976, and showing that for background rejection about $\sim$50, the signal efficiency is $\sim$75\%. Naturally, lower signal efficiency results in even better background rejection; for example, further increase of the background rejection by a factor of $\sim$100 will result in a signal efficiency reduction by a factor close to three.

\begin{figure}
\centering
\subfigure[]{\centering\includegraphics[width=0.45\textwidth]{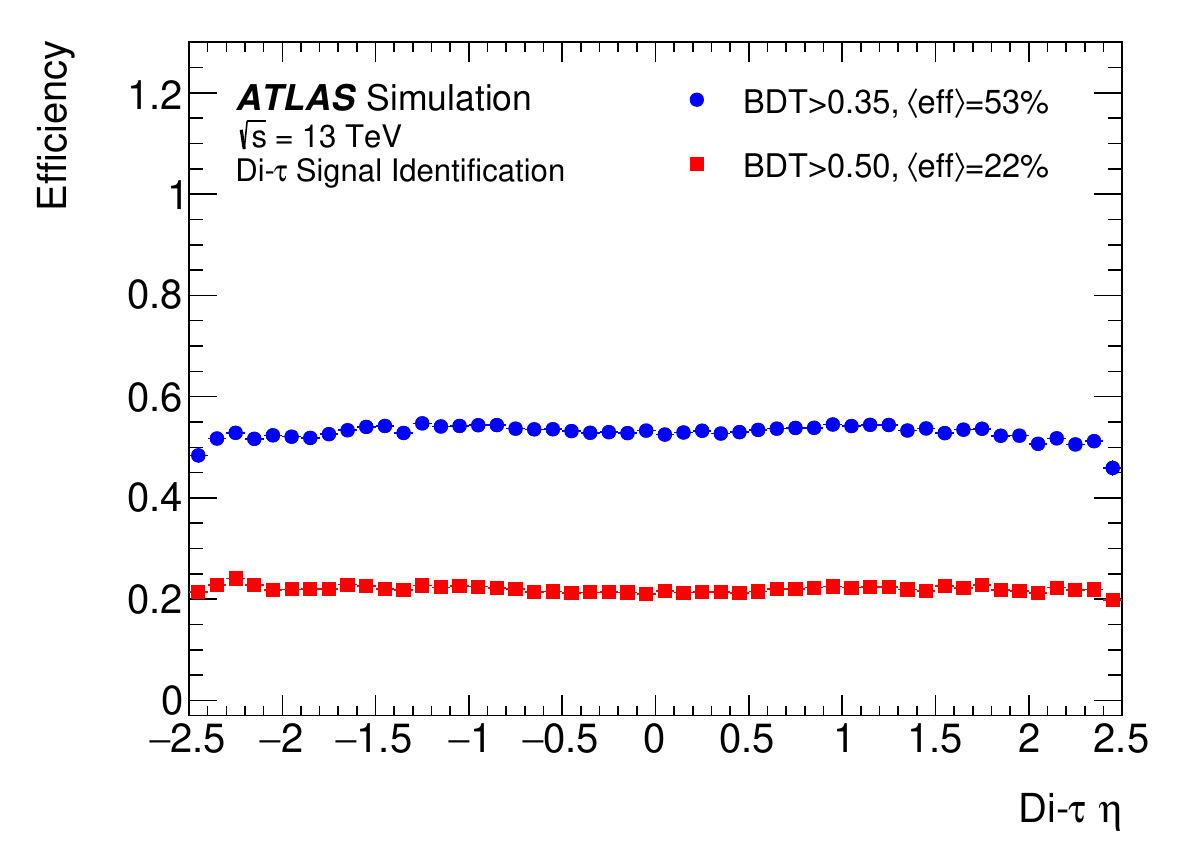}}
\subfigure[]{\centering\includegraphics[width=0.45\textwidth]{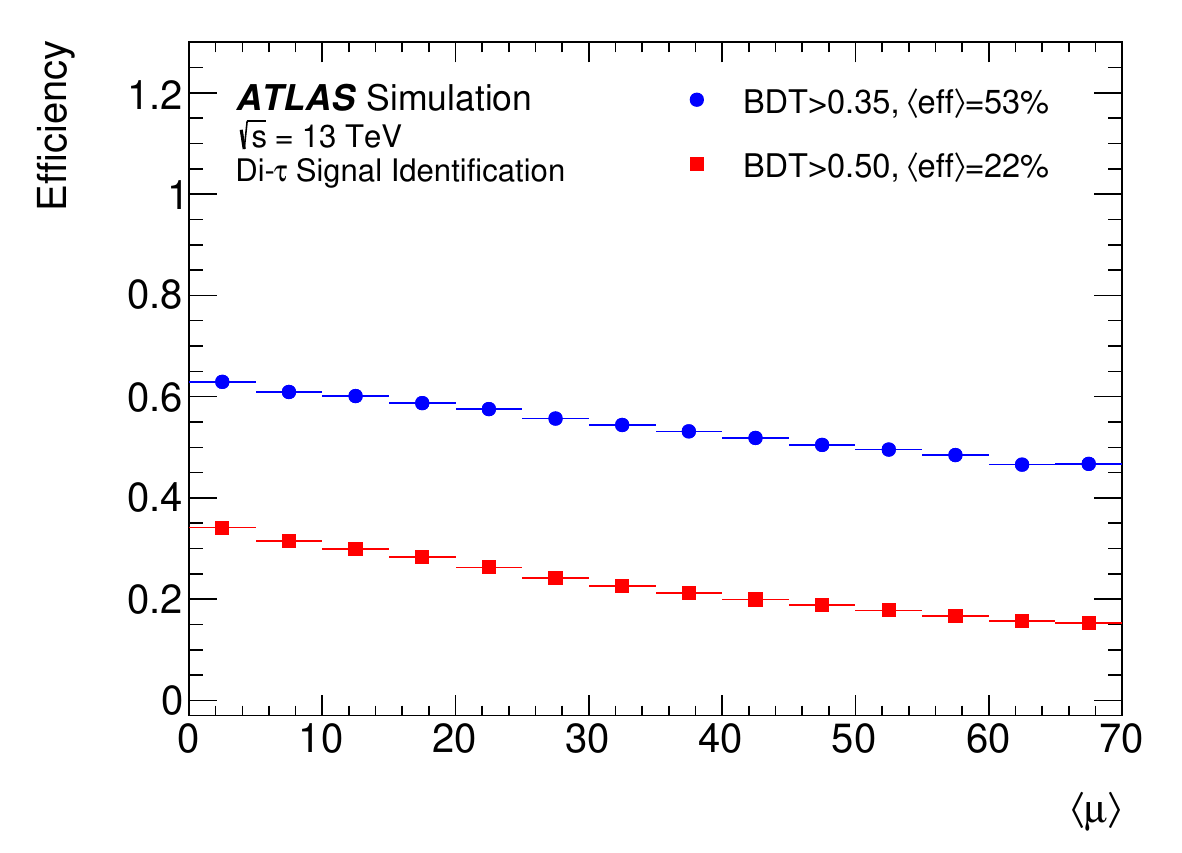}}
\subfigure[]{\centering\includegraphics[width=0.45\textwidth]{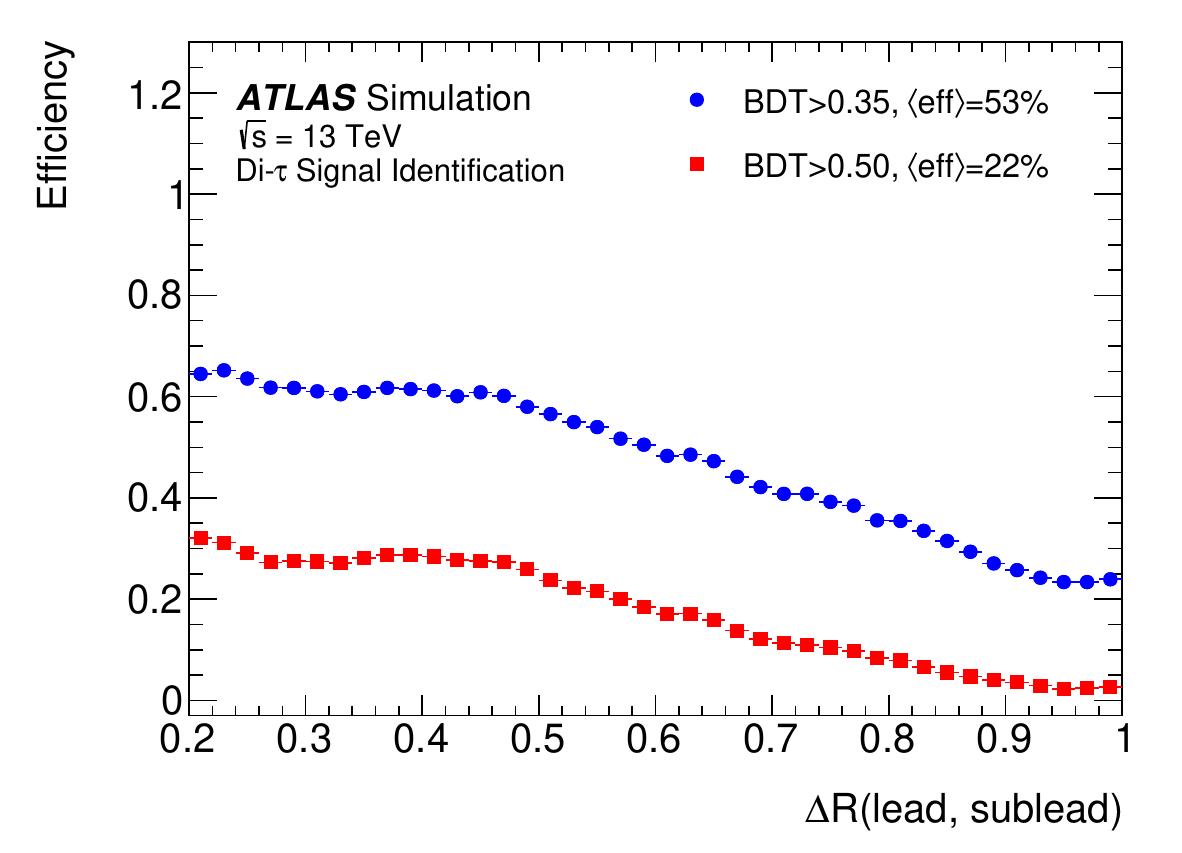}}
\subfigure[]{\centering\includegraphics[width=0.45\textwidth]{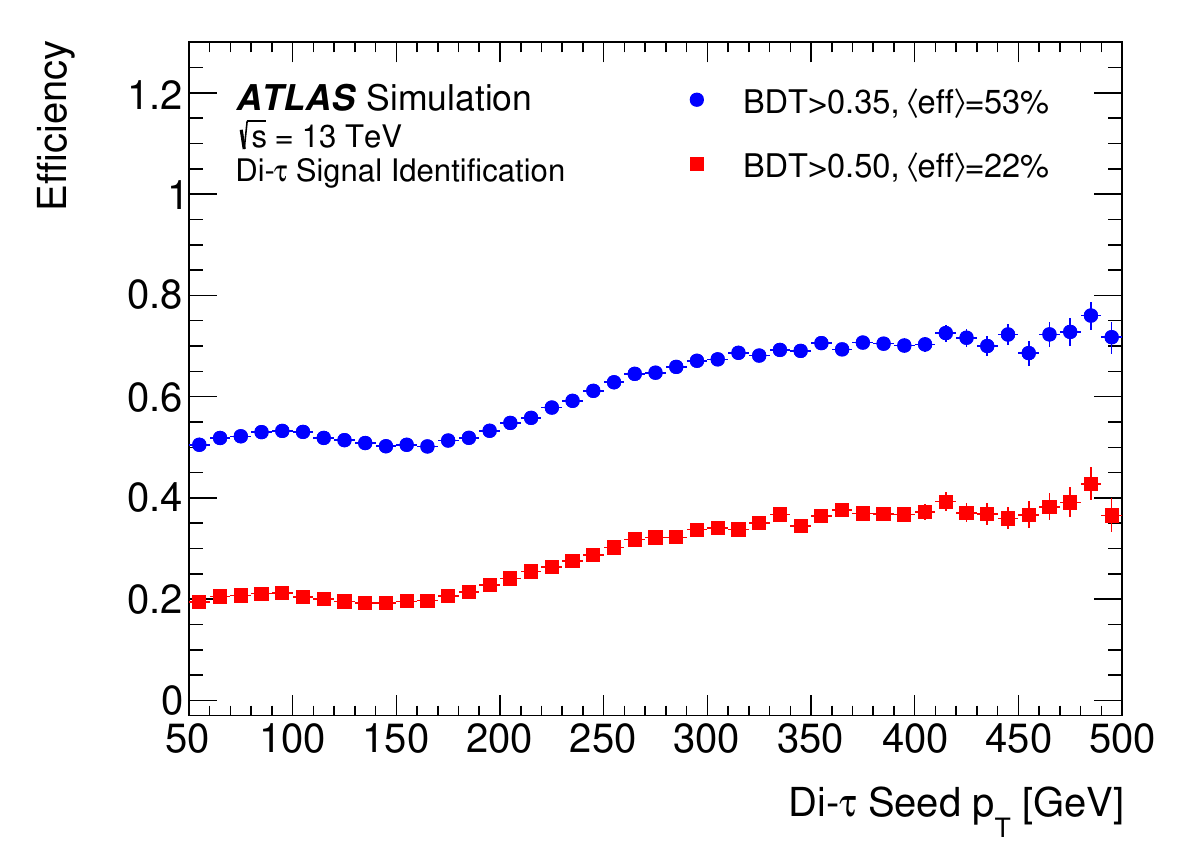}}
\caption{Signal identification efficiency at constant BDT selections, measured in $t\bar{t}X$ events, as functions of (a) the di-$\tau$ pseudorapidity $\eta_{\text{reco}}$, (b) the average number of interactions $\left<\mu\right>$, (c) the angular distance $\Delta R$ between the two leading di-$\tau$ subjets and (d) the transverse momentum of the di-$\tau$ seed jet. The error bars account for the statistical uncertainty in simulation. The $t\Bar{t}X$ sample includes events with $m_X$ = 20~\GeV and $m_X$ = 60~\GeV in equal proportions.
\label{fig:ideffs}}
\end{figure}

\begin{figure}
\centering
\subfigure[]{\centering\includegraphics[width=0.45\textwidth]{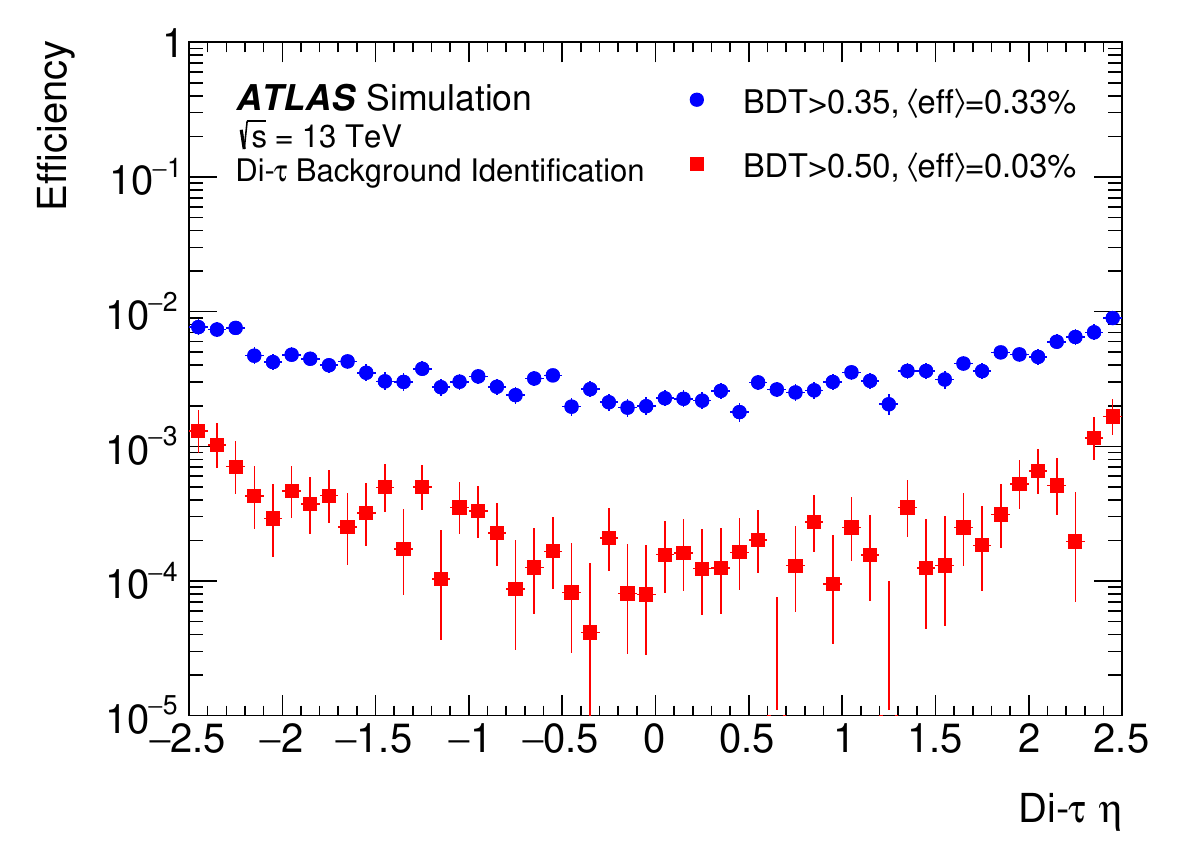}}
\subfigure[]{\centering\includegraphics[width=0.45\textwidth]{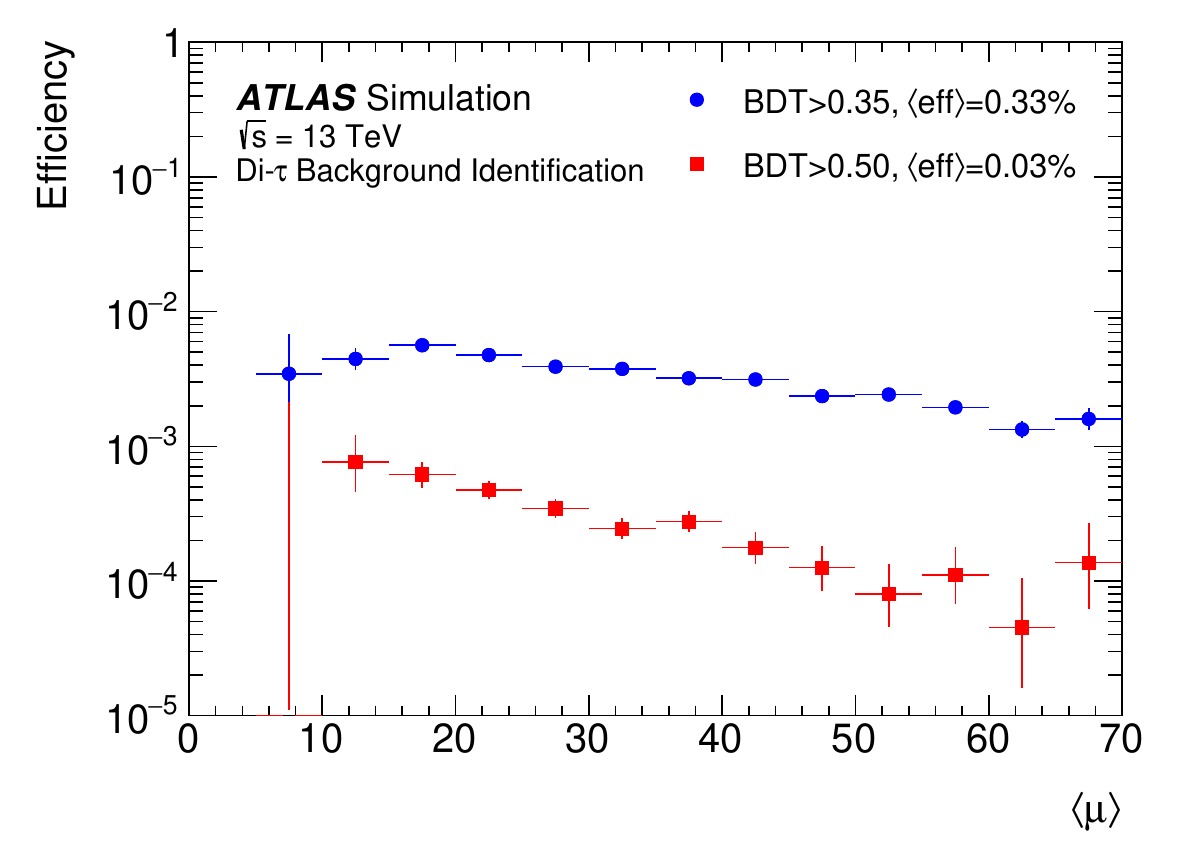}}
\subfigure[]{\centering\includegraphics[width=0.45\textwidth]{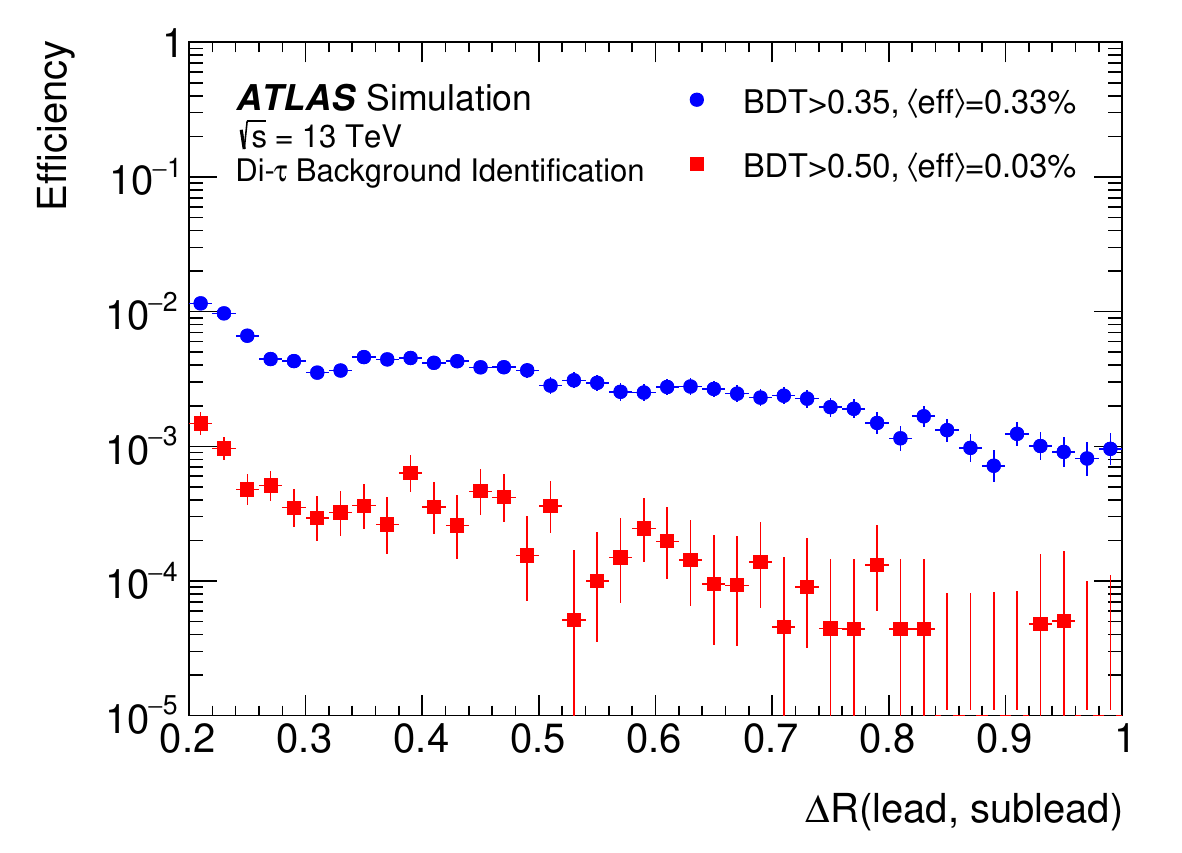}}
\subfigure[]{\centering\includegraphics[width=0.45\textwidth]{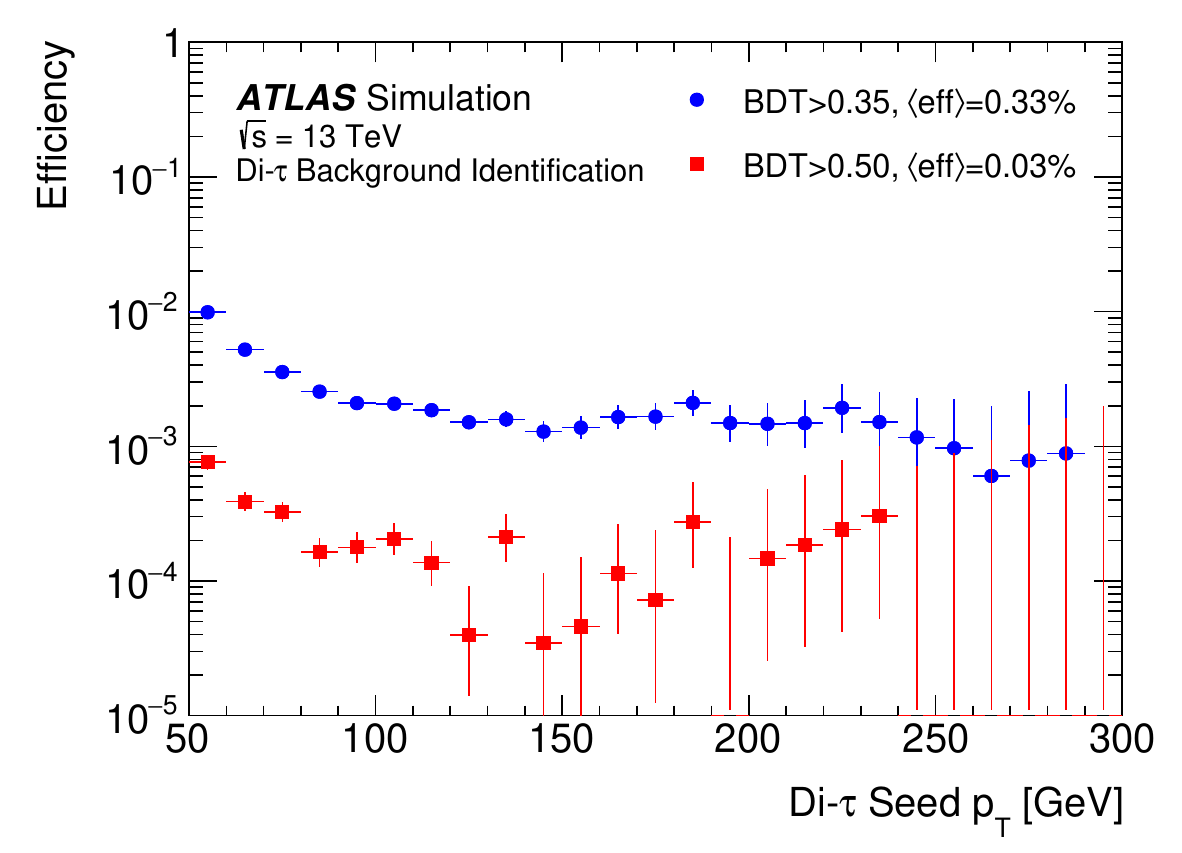}}
\caption{Background identification efficiency at constant BDT selections, measured in $t\bar{t}$ events, as functions of (a) the di-$\tau$ pseudorapidity $\eta_{\text{reco}}$, (b) the average number of interactions $\left<\mu\right>$, (c) the angular distance $\Delta R$ between the two leading di-$\tau$ subjets and (d) the transverse momentum of the di-$\tau$ seed jet. The error bars account for the statistical uncertainty in simulation.
\label{fig:id-bkg-rej}}
\end{figure}

Two benchmark WPs, labelled as \textit{Medium} and \textit{Tight}, are defined using constant selections on the classifier scores of 0.35 and 0.5, respectively.
The corresponding signal efficiencies are approximately 53\% and 22\% for rejection factors of 300 and 4000, respectively.
The dependence of the signal and background identification efficiencies for these WPs on several kinematic variables are illustrated in Figures~\ref{fig:ideffs} and~\ref{fig:id-bkg-rej}, respectively. The identification efficiency is approximately constant relative to the di-$\tau$ pseudorapidity $\eta_{\text{reco}}$, but clearly decreases as the average number of interactions $\left<\mu\right>$\ grows, a dependency which becomes more severe as the WP tightness increases. However, a similar behaviour is also observed for the background, such that the background rejection power increases as $\left<\mu\right>$\ increases. In terms of the angular separation between the two leading subjets, used as one of the input variables to the classifier, a decline in the identification efficiency as $\Delta R$ increases should be expected and is observed, with the classifier achieving its peak performance in the highly-boosted regime where the reconstruction is also most efficient. As the angular separation is inversely proportional to the transverse momentum at fixed $m_X$, the identification efficiency increases as the transverse momentum of the di-$\tau$ seed jet increases.

Signal and background BDT distributions are found to shift to lower values as the number of prongs in a subjet increases from one to three, leading to a decrease in both the signal and background efficiencies for a constant BDT score selection. This effect results in approximately 20\% efficiency loss per subjet in signal, and approximately a tenfold increase in rejection per subjet in background. As the shift is more significant for background than for signal, the overall performance of the identification improves with increasing subjet prongness -- in accordance with the performance observed for the resolved \tauhadvis identification~\cite{ATL-PHYS-PUB-2022-044}. Each three-pronged subjet improves the rejection by approximately a factor of two over its one-pronged counterpart, for a given constant signal efficiency.


%
%
%
%

%
%
\section{Identification efficiency measurement in $Z(\to\tau\tau)+\gamma$ events}
\label{sec:validation}
A dedicated tag-and-probe measurement is performed to measure the di-$\tau$ identification efficiency using $Z(\to\tau\tau)+\gamma$ signal events.
The identification efficiencies are obtained from data and simulated events, and their ratio is defined as the SF, computed for the previously noted \textit{Medium} and \textit{Tight} BDT-based identification WPs.
Few SM processes can be used to assess the performance of the di-$\tau$ identification algorithm in data. The $Z\gamma$ process was chosen due to its sizeable production cross section, the limited SM background contribution in the $\tau\tau\gamma$ final state, and the $Z$ boson \pt regime that can be accessed with a photon trigger. However, as the $Z$ boson mass is larger than the considered $m_X$ values, the phase space probed in the tag-and-probe measurement is not expected to match that of $t\Bar{t}X$ processes.
This measurement is statistically limited and hence performed inclusively in all di-$\tau$ kinematic properties.
\subsection{Event selection and categorisation}
Events are selected using the lowest unprescaled trigger requiring
the presence of a photon with $\et>140$~\GeV~\cite{TRIG-2018-05}.
The photon trigger-matching algorithm, which confirms the association between the reconstructed photon and the triggering signal, is applied.
Events containing electrons or muons are vetoed.

At preselection level, events are required to contain at least one photon passing the \textit{Tight} identification
and isolation working points~\cite{EGAM-2018-01} with $\pt>150$~\GeV and $\left|\eta\right|<2.37$.
Events with one or more photon candidates passing looser identification and isolation requirements
are also kept for  estimate of the backgrounds. Additionally, events are
required to contain at least one di-$\tau$ object,
which must satisfy the following requirements:
\begin{itemize}
\item The number of subjets is at least two.%
\item The invariant mass of the di-$\tau$ object is within the range $40<m_{\tau\tau}<130\text{~\GeV}$.
\item Each of the two leading subjets contains either one track (1-prong)
or three tracks (3-prong).
\item The charge product of the two leading subjets is $Q \equiv q_{\text{lead}}\times q_{\text{sublead}}=\pm1,$
where $q_{\text{lead}}$ ($q_{\text{sublead}}$) is the charge of
the (sub)leading subjet, defined as the sum of the charges of the
associated tracks.
\item The transverse momentum of the seed jet selection is optimised to $90<\pt^{\text{seed}}<360$~\GeV for achieving further background rejection. %
\end{itemize}
The leading $\pt$ photon and the leading BDT di-$\tau$ candidate are selected, and their angular separation is required to satisfy $\Delta R\left(\text{di-}\tau, \gamma\right)>1.0$,
selecting events where the di-$\tau$ object is well separated from the photon.
This requirement enhances the contribution of events with photons originating from initial-state radiation.

Events are further divided into control region (CR), validation region (VR) and signal region (SR), according to the charge product $Q$ of
the two leading di-$\tau$ subjets, and the azimuthal angle separation between the di-$\tau$ and \MET:
\begin{itemize}
\item CR: $Q=+1$, same-sign charges (SS).
\item VR: $Q=-1$, opposite-sign charges (OS); and $\Delta\phi\left(\text{di-}\tau, \MET\right)>2.2$.
\item SR: $Q=-1$, opposite-sign charges (OS); and $\Delta\phi\left(\text{di-}\tau, \MET\right)<2.2$.
\end{itemize}
In every region, if more than a single di-$\tau$ object fulfills the  requirements,
the one with the highest BDT score is selected.

Events from $Z\gamma$, $Z\gamma\gamma$ and $Z$+jets simulated samples, where the $Z$ boson decays into a $\tau^{+}\tau^{-}$ pair,
are classified as signal or background events based on whether the reconstructed boosted di-$\tau$
object passes the truth-matching requirement (indicating a real di-$\tau$, as defined in Section~\ref{sec:reco})
or not (indicating a misidentification), accordingly. Events from
all other samples are classified as background.

\subsection{Background estimation}

The dominant backgrounds arise from processes containing a real photon that satisfies the trigger requirement, together with a quark- or gluon-initiated jet that is misidentified as a di-$\tau$ object. Another source is processes containing multiple energetic jets, one being misidentified as a photon and another as a di-$\tau$ object.
These two sources correspond to $\gamma$+jets and multijet events respectively, and as QCD processes they are challenging to simulate. The $\gamma$+jets sample has as large as 20\% theory uncertainty in its cross-section~\cite{STDM-2017-01}, compared with only 5\% for the other samples~\cite{ATLAS-CONF-2015-039}. Therefore, a data-driven method is used to estimate their contribution, extrapolating the amount of background in the signal region from the yields observed in control regions.

The general strategy is as follows. A data-driven background estimate method is used to obtain the $\gamma$+jets and multijet normalisations, while distribution shapes are obtained from simulation.
To correct for the mismodelling of the BDT score distribution shape, events from $\gamma$+jets and multijet samples are reweighted to match the data according to a function that is obtained from the CR and tested in the VR.
The contribution of all other background components is taken from simulation, and normalised to their respective theoretical predictions.

\subsubsection*{Background normalisation}

The contribution from the $\gamma$+jets and multijet processes is estimated by using a sideband counting method~\cite{STDM-2012-16}, also referred to as the ABCD method.
This method relies on counting events with photon candidates in four regions of a two-dimensional plane, defined by the photon isolation and identification criteria.
A prompt photon region (region~A) is defined by photon candidates that are isolated and satisfy the \textit{Tight} identification, as explained in Section~\ref{sec:reconstruction}.
Three fake-photon regions are defined in the isolation-identification plane,
consisting of photon candidates that are non-\textit{Tight} and isolated (region B), \textit{Tight} and non-isolated (region C), or non-\textit{Tight} and non-isolated (region D).
A non-isolated photon candidate is defined by inverting the isolation requirement.
A photon candidate is classified as non-\textit{Tight} if it fails at least one of four selections associated with the shower-shape variables but passes all the other selections of the \textit{Tight} identification~\cite{EGAM-2018-01}.
Assuming negligible correlation between the photon identification and isolation would imply that the number of events with non-prompt photon candidates in the four regions ($N_{A},N_{B},N_{C}$ and $N_{D}$) satisfy the condition $N_{A}/N_{B}=N_{C}/N_{D}$, particularly considering the multijet sample.
The residual correlation between the photon isolation and identification is accounted for using the correlation factors defined as $R_{c}\equiv\left(N_{A}/N_{B}\right)/\left(N_{C}/N_{D}\right)$, and estimated from the multijet MC simulation.
A further correction is included to take into account the contamination from real photons in the three regions (B, C, or D) that are supposed to be dominated by fake photons.
This contribution is evaluated using the $\gamma$+jets MC simulation and is parameterised through the \textit{leakage coefficients}, representing the number of real photons in each of the aforementioned regions relative to the number of real photons in region A, i.e. $f_{\alpha} \equiv N^{\gamma}_{\alpha} / N^{\gamma}_{A}$, with $\alpha$ indicating region B, C, or D. Values for these two corrections are provided in Table~\ref{tab:ABCD-input}.
While MC simulations are used to estimate the leakage coefficients and correlation factor, different generator combinations for both the $\gamma$+jets and multijet samples are considered and corresponding uncertainties on these quantities are assigned.

In the ABCD method, the total SM prediction is normalised to data in each of the four regions, with the contribution from real photons and jets misidentified as photons, the $\gamma$+jets and multijets samples respectively, being unknown.
By assumption, the ABCD normalisation preserves the ratio represented by the leakage coefficients. Therefore, the $\gamma$+jets contribution in regions B, C, and D can be expressed via its normalised contribution in region~A.

The method determines the contribution of real photons $N_{A}^{\gamma}$ in region~A using the relation $N_{A}^{\text{MJ}}/N_{B}^{\text{MJ}} = R_{c} \cdot N_{C}^{\text{MJ}}/N_{D}^{\text{MJ}}$, with $N_{\alpha}^{\text{MJ}}$ being the estimated multijet contribution in region $\alpha$.
Along with the previously mentioned constraints, a single equation is obtained,
\begin{equation*}
\frac{N^{\text{data}}_{\text{A}}-N^{\text{MC}}_{\text{A}}-N^{\text{\ensuremath{\gamma}}}_{\text{A}}}{N^{\text{data}}_{\text{B}}-N^{\text{MC}}_{\text{B}}-f_{\text{B}}N^{\text{\ensuremath{\gamma}}}_{\text{A}}}
=
R_{c} \cdot \frac{N^{\text{data}}_{\text{C}}-N^{\text{MC}}_{\text{C}}-f_{\text{C}}N^{\text{\ensuremath{\gamma}}}_{\text{A}}}{N^{\text{data}}_{\text{D}}-N^{\text{MC}}_{\text{D}}-f_{\text{D}}N^{\text{\ensuremath{\gamma}}}_{\text{A}}}\ ,
\end{equation*}
where $N_{\alpha}^{\text{data}}$ and $N_{\alpha}^{\text{MC}}$ refer to the number of events in data and all signal and background MC samples excluding $\gamma$+jets and multijets events, respectively, in region~$\alpha$.

\begin{table}
\caption[Input to the ABCD method.]{Input to the ABCD method in the form of the leakage coefficients $f_B,f_C,f_D$ given in percentages and the correlation factor $R_c$, obtained from $\gamma$+jets and multijet MC samples, respectively, for the CR, VR, and SR.
Statistical uncertainties are given.
\label{tab:ABCD-input}}
\noindent \begin{centering}
\begin{tabular}{ccccc}
\hline
\hline
\multicolumn{2}{c}{} & \textbf{CR} & \textbf{VR} & \textbf{SR} \tabularnewline
\hline
\multirow{3}{*}{Leakage coefficients [\%]} & $f_B$ & 3.96 $\pm$ 0.32 & 2.83 $\pm$ 0.54 & 3.42 $\pm$ 0.27 \tabularnewline
& $f_C$ & 11.52 $\pm$ 0.52 & 9.3 $\pm$ 1.1 & 13.56 $\pm$ 0.61 \tabularnewline
& $f_D$ & 0.61 $\pm$ 0.13 & 0.67 $\pm$ 0.37 & 0.60 $\pm$ 0.10 \tabularnewline
\hline
\multicolumn{2}{c}{Correlation factor $R_c$} & 1.18 $\pm$ 0.89 & $4.3^{+5.1}_{-4.3}$ & 0.57 $\pm$ 0.38 \tabularnewline
\hline
\hline
\end{tabular}
\par\end{centering}
\vspace{10pt} %
\end{table}

Due to lack of statistical precision, the ABCD method is performed inclusively in each of the analysis regions, resulting in total yields predictions for $\gamma$+jets and multijet background samples in region A.
The fractions relative to data are summarised in Table \ref{tab:ABCD-results}, according to which in the SR, 0.1\% of data consists of multijet background, and nearly 96\% corresponds to prompt single photon production. Similar fractions for the real photons component are obtained for the CR and VR. This composition is indeed expected, as region A requires photons to satisfy the \textit{Tight} requirements for both identification and isolation, aiming to ensure the presence of prompt photons; those naturally exist at particle-level in the $\gamma$+jets sample, but are suppressed in the multijet sample.

\begin{table}
\caption[Background estimate results obtained using the ABCD method for the CR, VR, and SR.]{Background estimate results in the form of ABCD-estimated $\gamma$+jets and multijet contributions relative to data, for the CR, VR and SR.
Statistical uncertainties are given.
\label{tab:ABCD-results}}
\noindent \begin{centering}
\begin{tabular}{cccccc}
\hline
\hline
& \textbf{CR} & \textbf{VR} & \textbf{SR}\tabularnewline
\hline{$\gamma$+jets fraction [\%]} & {94.8 $\pm$ 1.6} & {91 $\pm$ 10} & {96.0 $\pm$ 1.2} \tabularnewline
{Multijet fraction [\%]} & {1.2 $\pm$ 1.1} & {$4.3^{+9.7}_{-4.3}$} & {$0.09^{+0.29}_{-0.09}$} \tabularnewline
\hline
\hline
\end{tabular}
\par\end{centering}
\vspace{10pt} %
\end{table}

Furthermore, the multijet sample shows significant statistical loss upon the application of the preselection criteria. To ensure a smooth distribution of the di-$\tau$ BDT score, all photon selections are omitted in this simulated sample, retaining only the di-$\tau$ $\pt$ selection, while the expected multijet background yield is estimated from data events and not directly obtained from the MC sample.
The di-$\tau$ BDT score distribution was compared for different photon selections applied, and found to only weakly depend on selections associated with the photon.

\begin{figure}
\begin{centering}
\includegraphics[width=0.6\textwidth]{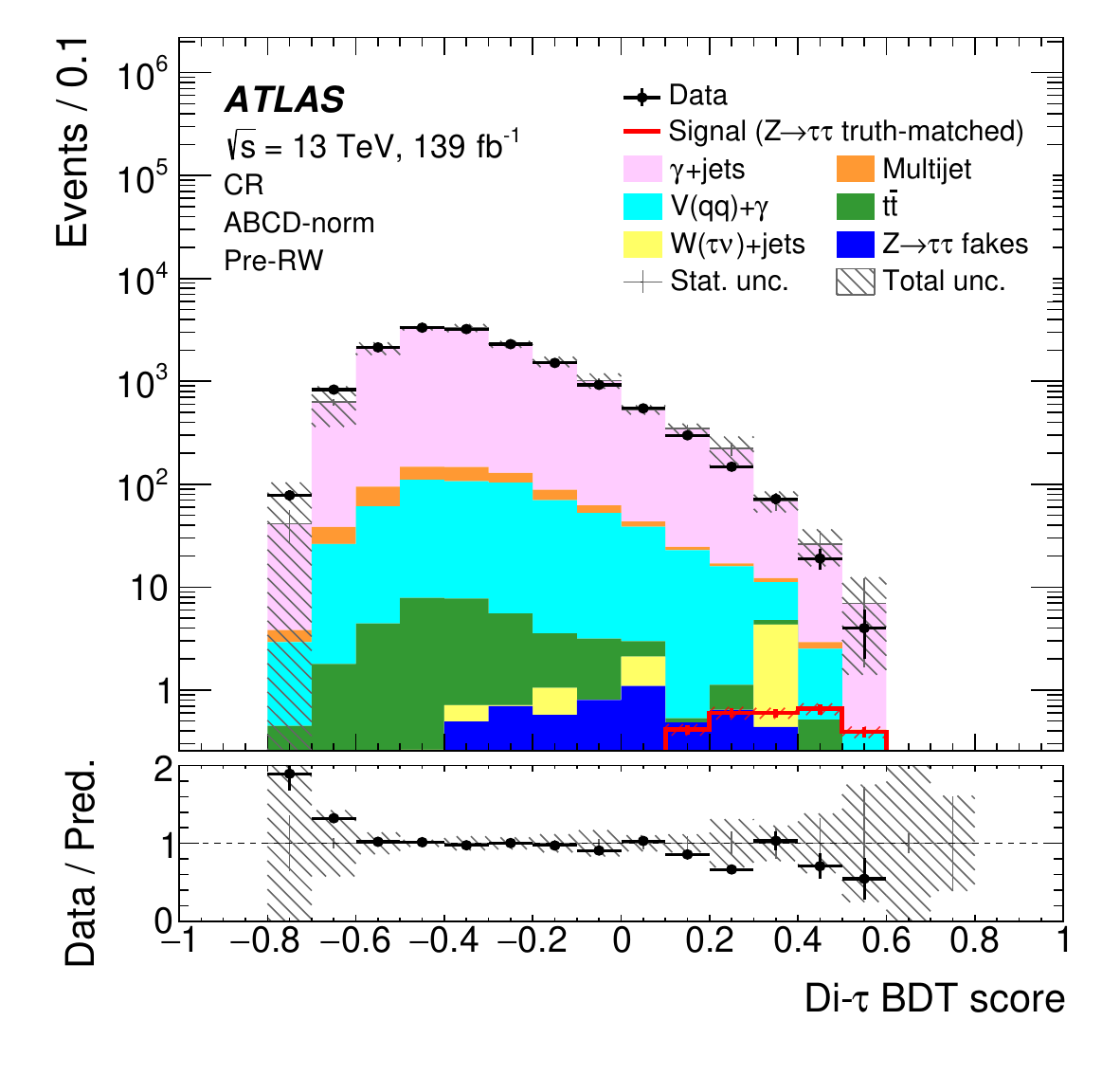}
\caption{Di-$\tau$ BDT score distribution with $\gamma$+jets and multijet yields ABCD-normalised in the CR, before event reweighting.
Simulated events from $Z\gamma$, $Z\gamma\gamma$ and $Z$+jets, where $Z\rightarrow\tau\tau$, containing a reconstructed di-$\tau$ object geometrically matched to a particle-level $\tau^{+}\tau^{-}$ pair are referred to as Signal (red open histogram). The lower panel shows a bin-by-bin comparison of the data with the total predicted background and signal, in terms of event yield ratios.
The uncertainty band corresponds to both the statistical and systematic uncertainties in the total SM prediction, detailed in Section~\ref{sec:sys}. Error bars on the markers represent the statistical uncertainty in data.
\label{fig:Ditau-BDT-before-reweighting}}
\end{centering}
\end{figure}

The ABCD-normalised di-$\tau$ BDT score distribution for the CR is shown in Figure~\ref{fig:Ditau-BDT-before-reweighting}. A trend can be seen in the data-to-total prediction ratio, implying a mismodelling of the di-$\tau$ BDT in the $\gamma$+jets sample. Since the ABCD method does not have any impact on the shape, a reweighting approach was developed to correct for this mismodelling.

\subsubsection*{Event reweighting}

The dominant background in the analysis is $\gamma$+jets, accounting for over 95\% of pre-selected data events, and therefore is expected to resemble the distribution for data.
However, as seen in Figure~\ref{fig:Ditau-BDT-before-reweighting}, the di-$\tau$ BDT score distribution shape for $\gamma$+jets shows a discrepancy as compared with the data, and applying event reweighting can result in improved modelling.
To mitigate a bias in the SF measurement arising from signal contribution in the reweighting process, a reweighting factor is derived from the BDT score distribution in the CR. The signal significance in the CR compared with the SR is smaller by factors of 22, 29, and 51 for the full BDT range, BDT > 0.35 and BDT > 0.5, respectively, such that biases due to signal contamination are not expected.
Furthermore, since the primary distinction between multijet and $\gamma$+jets processes is due to the presence of a prompt photon, and since both contain an energetic jet misidentified as a di-$\tau$ object, the event reweighting is applied to both of the samples.
The reweighting factor, binned in BDT score values, is hence defined as $\left(N_{\text{data}}-N_{\text{MC bkg}}\right)\slash\left(N_{\gamma}+N_{\text{MJ}}\right)$, where $N_{\text{data}}$ is the number of events in the observed data, $N_{\text{MC bkg}}$ refers to number of events in all MC background processes except for $\gamma$+jets and multijet, and $N_{\gamma},N_{\text{MJ}}$ refer to the ABCD-estimated yield for the $\gamma$+jets and multijet processes, respectively.

\begin{figure}
\begin{centering}
\subfigure[]{\includegraphics[width=0.45\textwidth]{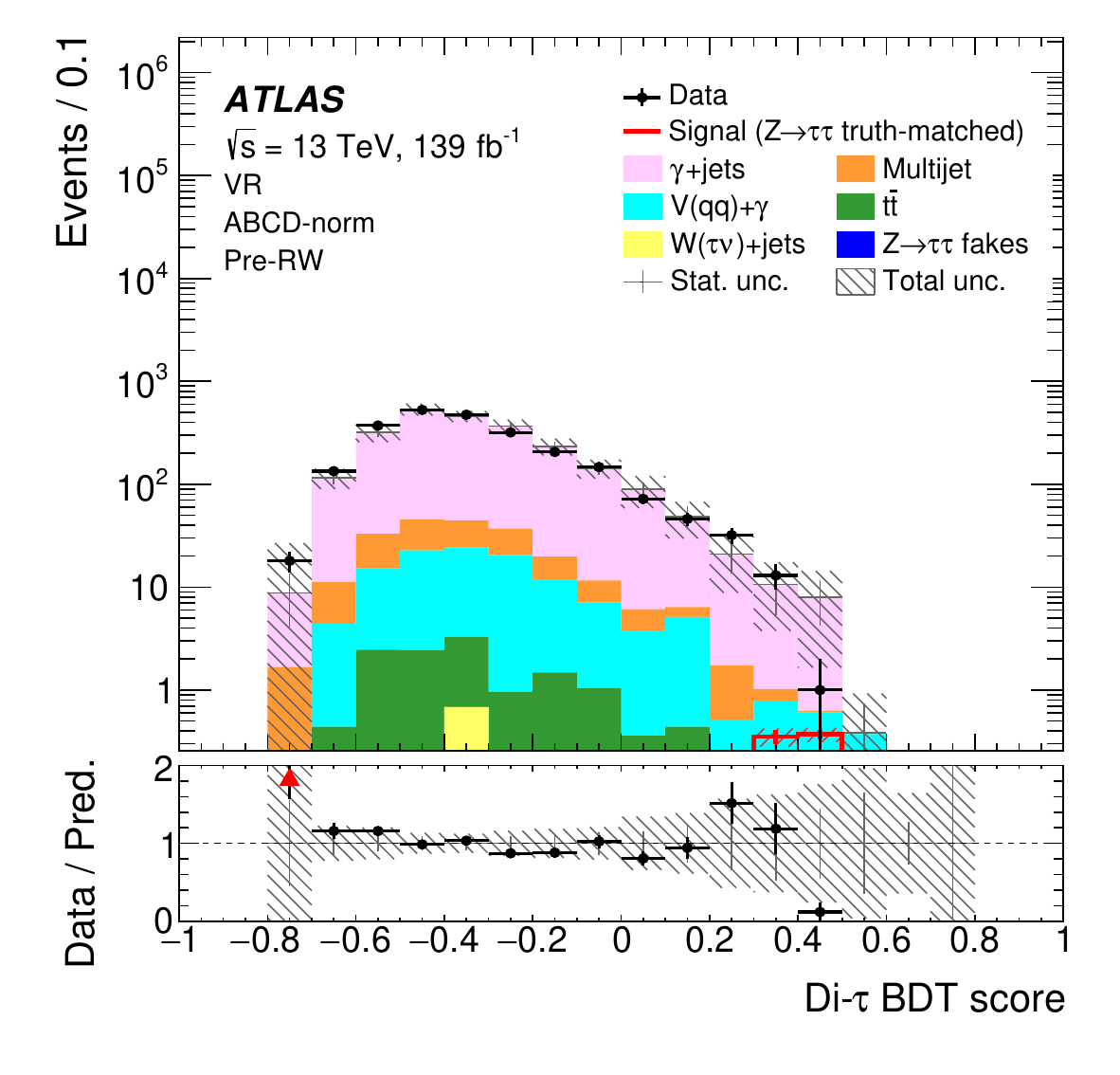}\label{fig:VRa}}
\subfigure[]{\includegraphics[width=0.45\textwidth]{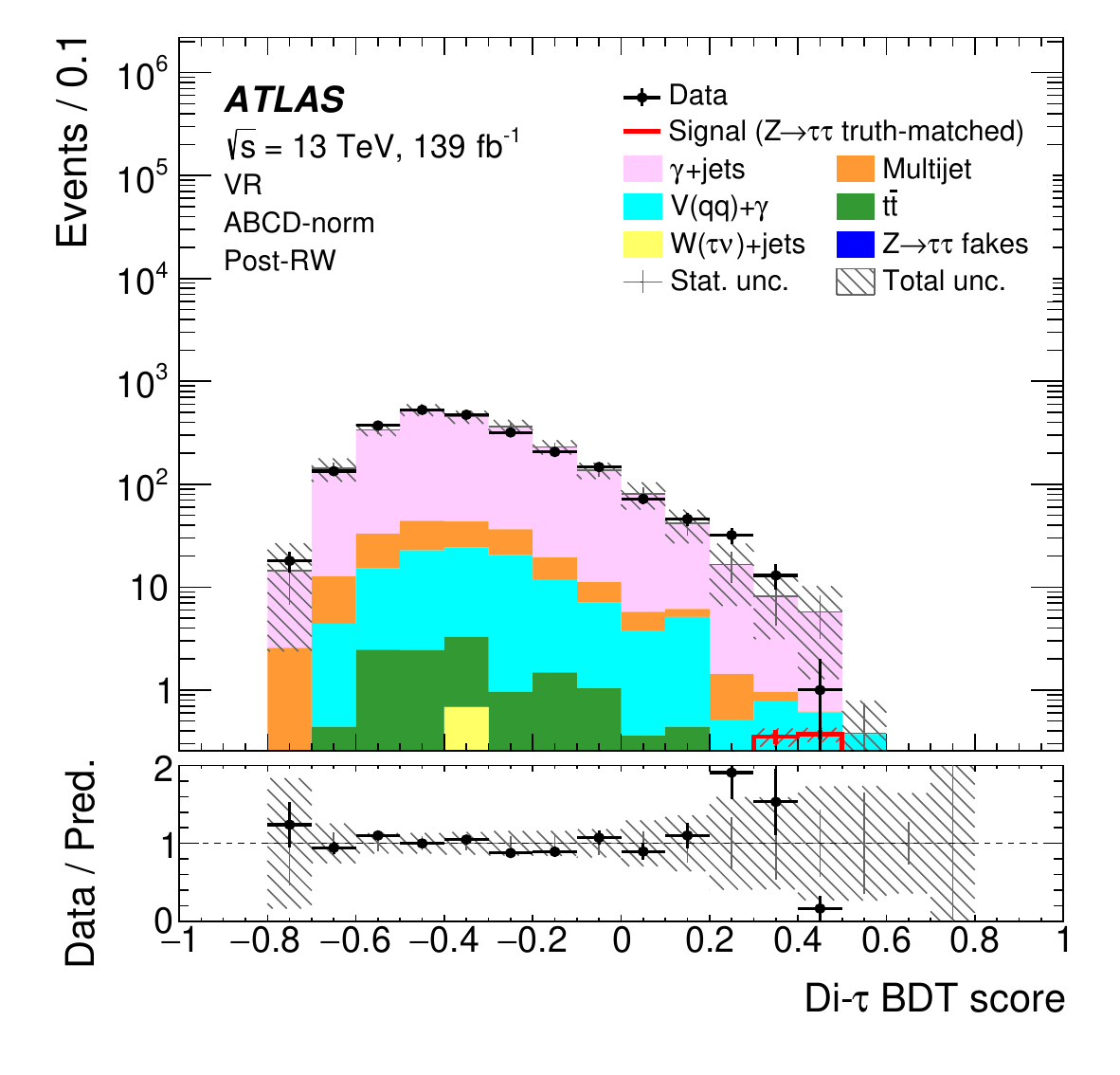}\label{fig:VRb}}

\subfigure[]{\includegraphics[width=0.45\textwidth]{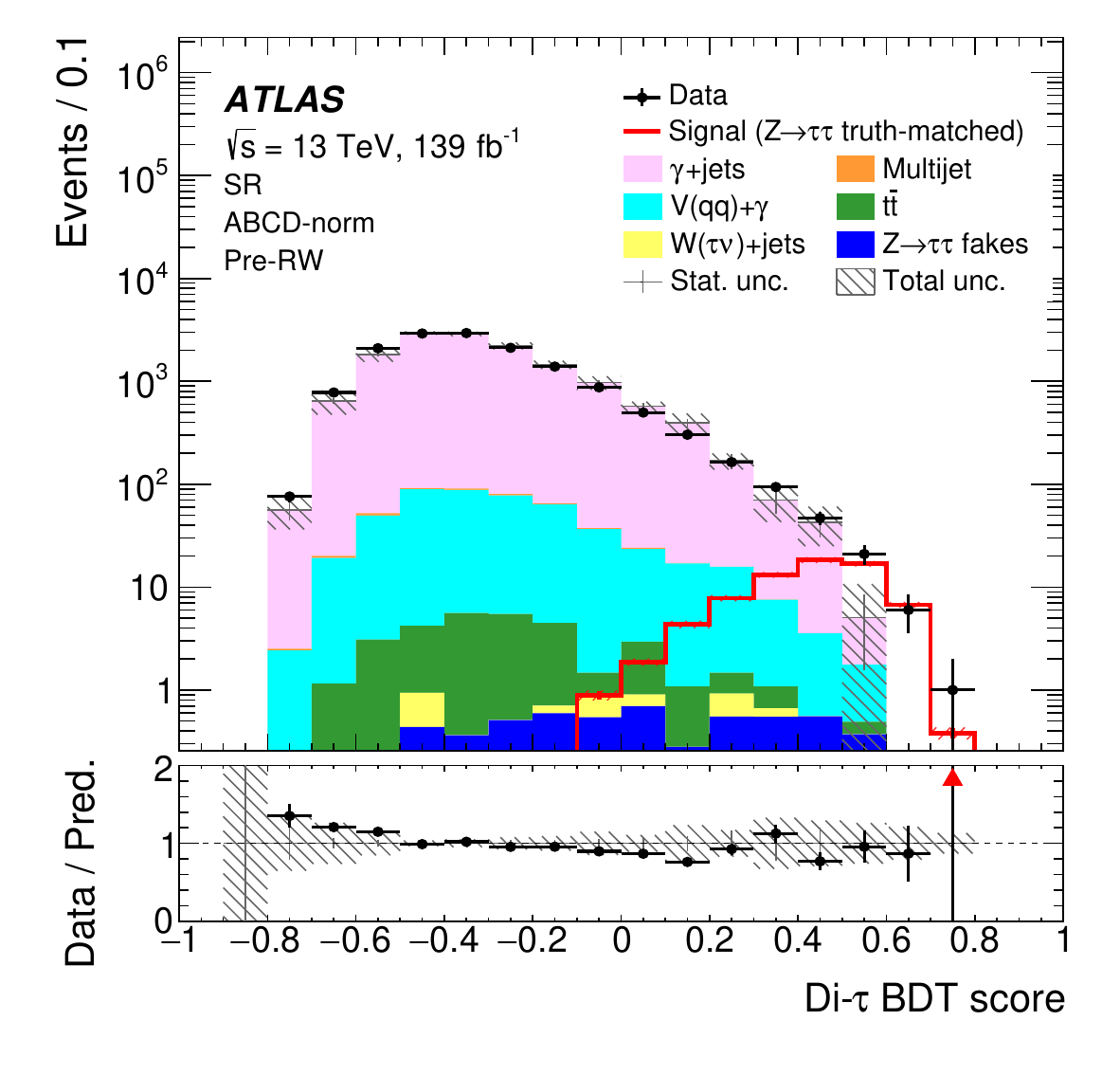}\label{fig:final-Ditau-BDTa}}
\subfigure[]{\includegraphics[width=0.45\textwidth]{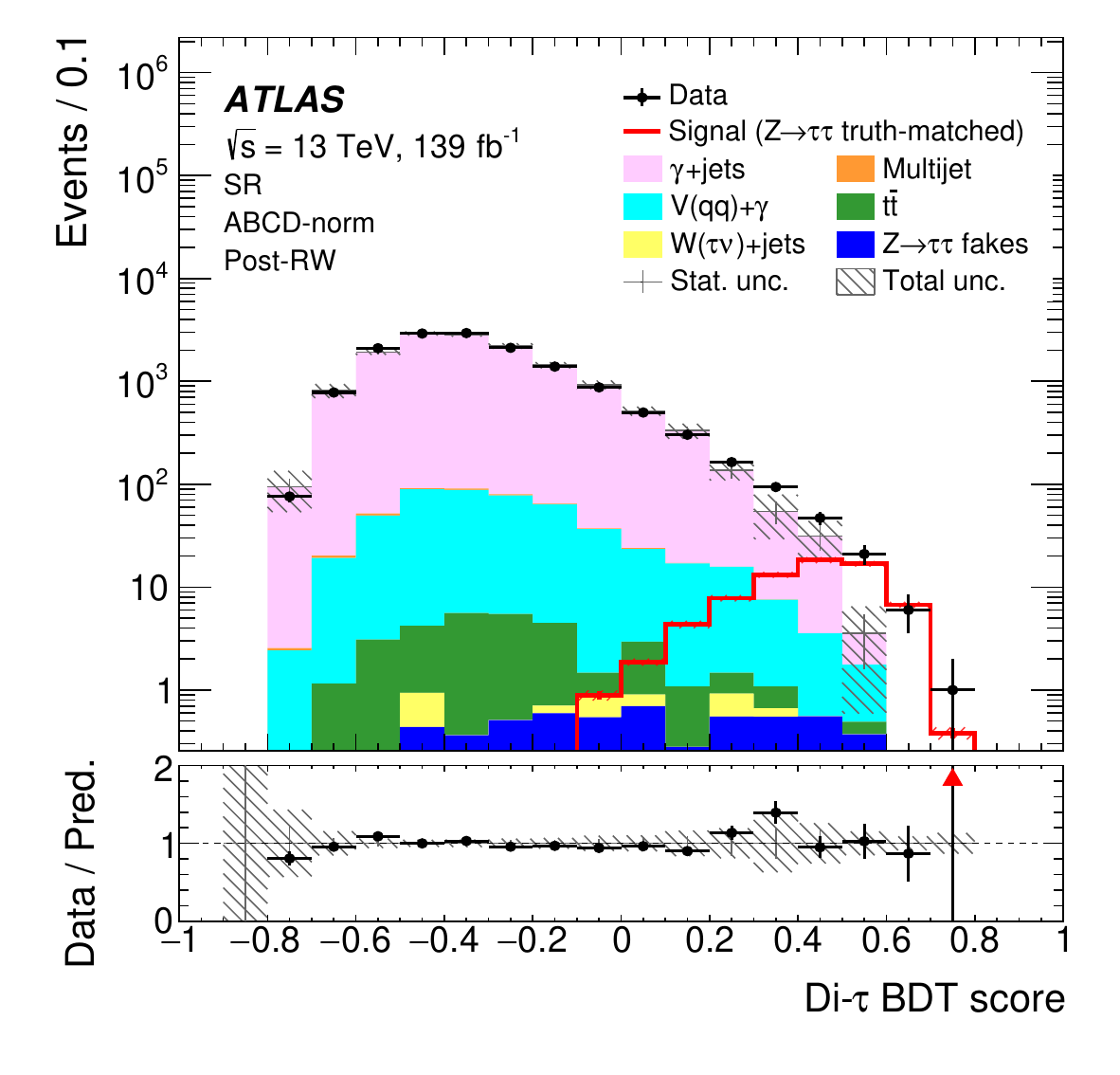}\label{fig:final-Ditau-BDTb}}

\caption{Di-$\tau$ BDT score distributions with $\gamma$+jets and multijet yields ABCD-normalised in the VR (a) before and (b) after event reweighting, and in the SR (c) before and (d) after event reweighting.
Simulated events from $Z\gamma$, $Z\gamma\gamma$ and $Z$+jets, where $Z\rightarrow\tau\tau$, containing a reconstructed di-$\tau$ object geometrically matched to a particle-level $\tau^{+}\tau^{-}$ pair are referred to as Signal (red open histogram).
The lower panel shows a bin-by-bin comparison of the data with the total predicted background and signal, in terms of event yield ratios.
Vertical arrows indicate points that lie outside the displayed axis range.
The uncertainty band corresponds to both the statistical and systematic uncertainties in the total SM prediction, detailed in Section~\ref{sec:sys}. Error bars on the markers represent the statistical uncertainty in data.
\label{fig:final-Ditau-BDT}}
\end{centering}
\end{figure}

The reweighting function that is eventually employed is an analytical function that is fitted to the reweighting factor, and as such is applied as a BDT score-dependent event weight.
Two criteria are considered for selecting the optimal function: the $\chi^{2}$ test, evaluating the goodness of the fit in the CR, and the agreement between the data and the total prediction after reweighting in the VR.
Various functions were tested on the VR, of which a fifth-order polynomial is chosen as the nominal reweighting function, while other functional forms are used to derive an associated uncertainty on the reweighting process.
The pre-reweighting di-$\tau$ BDT score distribution in the VR is shown in Figure~\ref{fig:VRa}, with the corresponding post-reweighting distribution given in Figure~\ref{fig:VRb}.
The latter shows the reweighting procedure impact on the VR -- with the reweighting uncertainty included in the uncertainty band -- increasing the total contribution from $\gamma$+jets and multijets samples by about 10\% for the BDT > 0.2 domain.

The reweighting is then applied to events from $\gamma$+jets and multijet samples in the SR, modifying their contributions by about $+12\%$ and $-1\%$, respectively, for the BDT > 0.2 domain.
The pre- and post-reweighting BDT score distributions are shown in Figure~\ref{fig:final-Ditau-BDT}\subref{fig:final-Ditau-BDTa}-\subref{fig:final-Ditau-BDTb}, with ABCD-estimated yields given in Table~\ref{tab:ABCD-results}.
The event reweighting primarily affects the shape, showing negligible impact on the total normalisation of the background samples obtained through the ABCD method before reweighting, with a difference of less than 0.5\%.
Although designed to achieve better modelling for the background in the CR and VR, where the signal contribution is negligible, upon comparing the data-to-total prediction ratio in the SR, the declining trend vanishes with event reweighting. This correction effectively resolves the observed discrepancy, such that the total prediction and data are now within the full uncertainty band.

The reweighted di-$\tau$ BDT score distribution is used for computing the SF, as described in Section~\ref{subsec:reults}.
\subsection{Systematic uncertainties}
\label{sec:sys}

The systematic uncertainties that affect the SF measurements are divided into four categories:
experimental uncertainties affecting the simulated background and signal processes,
uncertainties derived from using different generators for $\gamma$+jets and multijet samples,
uncertainties in the modelling of the reweighting factor,
and theoretical uncertainties of the simulated background and signal samples.
All systematic uncertainties are propagated through all the analysis chain, including the ABCD method (and particularly the leakage coefficients and correlation factor), being reprocessed independently for each systematic variation. Furthermore, fixing $R_{c}$ to one to assess the impact of correlations in the ABCD method, results with background fractions differing by less than 0.1\% in the SR, and hence a negligible effect on the SF.
Uncertainties are added up in quadrature to express the total systematic uncertainty that corresponds to the SF, with the dominant sources listed in Table~\ref{tab:unc}.

Experimental uncertainties address the luminosity determination and modelling of detector effects.
The leading effects come from those associated with the modelling of pile-up interactions in simulation, photon energy scale and resolution, muon and \met modelling, and di-$\tau$ detector modelling and energy-scale calibration.
The largest contribution for the \textit{Medium} WP stems from the pile-up reweighting due to the low statistical precision in the $\gamma$+jet sample in the corresponding BDT region.

The estimate of di-$\tau$ detector modelling and energy-scale calibration uncertainties uses four systematically varied $t\bar{t}X$ MC simulations.
The di-$\tau$ reconstruction efficiency is parameterised in the particle-level di-$\tau$ \pt and $\Delta R_{\text{vis}}$.
The differences in efficiency of each of the varied samples to the nominal sample are summed in quadrature and are assigned as the di-$\tau$ detector modelling uncertainty, which are applied to truth-matched di-$\tau$ objects.
The di-$\tau$ energy-scale uncertainty is parameterised in the calibrated di-$\tau$ \pt, $\Delta R(\text{lead,\ sublead})$ and the subjet prongness. The calibrated subjet \pt distribution from the nominal sample is compared with the varied ones via a scaling parameter using a $\chi^2$ test, and the relative uncertainty is taken as the value of the scaling parameter that minimises the $\chi^2$ distribution. The contributions from the four different varied simulations are summed in quadrature to obtain the total uncertainty, which is applied as up and down variations on subjets \pt of truth-matched di-$\tau$ objects.

Different generators for the $\gamma$+jets and multijet samples are used to estimate the parton-shower and hadronisation uncertainties. The choice for the nominal generators is \SHERPA[2.2.2] for $\gamma$+jets and \POWPY for multijet. Two varied combinations are obtained using the nominal generator for one of the samples together with either \PYTHIA for $\gamma$+jets or \POWHER for multijet.
These variations are particularly important in the determination of uncertainties in the simulation-derived correlation factor and leakage coefficients, used to obtain the background normalisation.
In both of the combinations, background events are reweighted using a dedicated fifth-order polynomial function obtained from the di-$\tau$ BDT distribution in their CR.
The uncertainty considers the largest deviation from the nominal value.

To account for small variations in the SF derived from different choices of the fitting function, a corresponding uncertainty is introduced. This uncertainty considers alternative reweighting functions, chosen for their similarity to the nominal function in terms of the reduced $\chi^{2}$ goodness-of-fit parameter. Specifically, a quartic polynomial and a sum of two tangent functions are selected. The uncertainty is determined as the largest deviation from the nominal value.

Uncertainties in the calculation of the cross-sections for the different
processes are considered. Following the recommendations given in Ref.~\cite{ATLAS-CONF-2015-039},
a 5\% total theoretical uncertainty for the estimate
of the expected event yields is assigned for all MC samples from
which the normalisation is utilised, which accounts for the uncertainties
arising from the choice of renormalisation and factorisation scales
and the PDF choice.
For the $\gamma$+jets sample, scale uncertainties are used instead, and evaluated by varying the renormalisation and factorisation scales, $\mu_{\text{R}}$ and $\mu_{\text{F}}$, independently by factors of two and one-half, removing combinations where the variations differ by a factor of four.

\begin{table}
\caption{Summary of the dominant systematic uncertainties in the SF. The uncertainties
are expressed as percentages relative to the nominal SF. Only
uncertainties exceeding 1\% are shown.\label{tab:unc}}
\begin{centering}
\begin{tabular}{llcc}
\hline
\hline
\multirow{2}{*}{\textbf{Category}} & \multirow{2}{*}{\textbf{Source}} & \multicolumn{2}{c}{\textbf{Relative uncertainty [\%]}}\tabularnewline
\cmidrule{3-4} \cmidrule{4-4}
&  & \textbf{\textit{Medium}} & \textbf{\textit{Tight}}\tabularnewline
\hline

\multirow{3}{*}{Statistical}
& Data & 21 & 22\tabularnewline
& MC & 28 & 8.3 \tabularnewline
& \textit{Total} & 35 & 23\tabularnewline
\hline

\multirow{4}{*}{Modelling}
& Generators & 3.1 & 9.3\tabularnewline
& Reweighting & 3.7 & 4.1\tabularnewline
& Theory & 6.4 & 5.6\tabularnewline
& \textit{Total} & 8.0 & 12\tabularnewline
\hline

\multirow{9}{*}{Experimental}
& Muon sagitta evaluation & --- & 1.1\tabularnewline
& \MET resolution & 1.0 & --- \tabularnewline
& Integrated luminosity & 1.6 & 1.6 \tabularnewline
& Di-$\tau$ detector modelling & 1.9 & 1.2 \tabularnewline
& Di-$\tau$ calibration & 1.9 & 1.7\tabularnewline
& Photon energy resolution & 2.8 & --- \tabularnewline
& Photon energy scale & 4.0 & --- \tabularnewline
& Pile-up reweighting & 8.7 & 2.0\tabularnewline
& \textit{Total} & 9.9 & 3.6\tabularnewline
\hline

\multicolumn{2}{l}{Total uncertainty} & 37 & 26\tabularnewline

\hline
\hline
\end{tabular}
\par\end{centering}
\vspace{10pt} %
\end{table}

\subsection{Results}
\label{subsec:reults}
The di-$\tau$ identification efficiency SF is computed in the SR as the ratio of observed to expected signal efficiencies, for a specified di-$\tau$ BDT score selection.
As the ABCD method fixes the total expected yield to data, the definition is reduced to a ratio of event yields.
Since the predicted BDT inclusive signal purity is less than 1\%, its impact on the ABCD-estimated background normalisation and therefore on the SF is negligible.
The measured signal yield is obtained after subtracting the backgrounds with a misidentified di-$\tau$ object from the data:
\begin{equation}
\text{SF}=\frac{N_{\text{data}}-N_{\text{non-di-$\tau$}}}{N_{\text{true di-$\tau$}}}\ .\label{eq:SF-1}
\end{equation}
The yields for data, background and signal samples are obtained by applying the desired di-$\tau$ BDT score selection in the corresponding distribution (Figure~\ref{fig:final-Ditau-BDTb}), and are given in Table \ref{tab:Total-yields}.
The resultant signal identification efficiency is approximately 70\% (34\%) at background rejection rate of approximately 240 (3600) for the \textit{Medium} (\textit{Tight}) WP.
These efficiencies differ from those estimated using the BDT training samples, shown in Figure~\ref{fig:ideffs}, indicating that the identification efficiency depends on the kinematic phase space in which the measurement is performed.

\begin{table}
\caption{Total yields for all relevant samples that are required for the SF calculation: BDT inclusive yields, followed by \textit{Medium} (BDT > 0.35) and \textit{Tight} (BDT > 0.5) WP yields with the corresponding BDT efficiencies.
'MC background' refers to background events from all MC samples excluding $\gamma$+jets and multijet samples.
'MC signal' refers to simulated events from $Z\gamma$, $Z\gamma\gamma$ and $Z$+jets, where $Z\rightarrow\tau\tau$, containing a truth-matched reconstructed di-$\tau$ object.
Events from both the $\gamma$+jets and multijet samples are ABCD-normalised and reweighted.
Total uncertainties are given.
\label{tab:Total-yields}}
\noindent \begin{centering}

\begin{tabular}{lccccc}
\hline
\hline
\multirow{2}{*}{\textbf{Sample}} & \multirow{2}{*}{\textbf{BDT inclusive}} & \multicolumn{2}{c}{\textbf{\textit{Medium}}} & \multicolumn{2}{c}{\textbf{\textit{Tight}}}\tabularnewline
\cmidrule{3-6} \cmidrule{4-6} \cmidrule{5-6} \cmidrule{6-6}
&  & \textbf{Yield} & \textbf{Efficiency [\%]} & \textbf{Yield} & \textbf{Efficiency [\%]}\tabularnewline
\hline{MC signal} & {71.0 $\pm$ 3.3} & {49.7 $\pm$ 3.0} & {70.0 $\pm$ 4.2} & {24.0 $\pm$ 1.4} & {33.8 $\pm$ 2.0}\tabularnewline{MC background} & {496 $\pm$ 23} & {7.8 $\pm$ 1.2} & {1.57 $\pm$ 0.24} & {1.94 $\pm$ 0.63} & {0.39 $\pm$ 0.13}\tabularnewline{$\gamma$+jets} & {13750 $\pm$ 370} & {52 $\pm$ 15} & {0.38 $\pm$ 0.11} & {1.8 $\pm$ 3.2} & {0.01 $\pm$ 0.02}\tabularnewline
{Multijet} & {$13^{+130}_{-13}$} & {$0.03^{+0.38}_{-0.03}$} & {$0.3^{+3.0}_{-0.3}$} & {$<0.01$} & {$<0.01$}\tabularnewline
\hline
Total predicted & {14330 $\pm$ 390} & {110 $\pm$ 15} & {0.77 $\pm$ 0.11} & {27.8 $\pm$ 3.6} & {0.19 $\pm$ 0.02}\tabularnewline{Data} & {14330} & {110} & {0.77} & {28} & {0.20}\tabularnewline
\hline
\hline
\end{tabular}{\footnotesize\par}
\par\end{centering}
\vspace{10pt} %
\end{table}

The SF is calculated for \textit{Medium} and \textit{Tight} BDT-based di-$\tau$ identification requirements, for which BDT score $>0.35$ and $>0.5$ selections are chosen, respectively.
Being a statistically-limited study, an inclusive SF is obtained.
From Eq. \eqref{eq:SF-1}, using the total yields given in Table \ref{tab:Total-yields}, %
the SFs are finally obtained:
\begin{equation*}
\text{SF}\left(\text{BDT}>0.35\right) = 1.00 \pm 0.35\left(\text{stat.}\right) \pm 0.13\left(\text{syst.}\right) = 1.00 \pm 0.37\left(\text{tot.}\right),  \nonumber%
\end{equation*}
and
\begin{equation*}
\text{SF}\left(\text{BDT}>0.5\right) = 1.01 \pm 0.24\left(\text{stat.}\right) \pm 0.12\left(\text{syst.}\right) = 1.01 \pm 0.27\left(\text{tot.}\right).  \nonumber %
\end{equation*}

The total relative uncertainty is about 37\% (26\%) for the \textit{Medium} (\textit{Tight}) WP.
The SFs are found to be compatible with 1, well within the associated uncertainties.
Notably, the statistical uncertainty for the \textit{Medium} WP is larger than for the \textit{Tight} WP. This counterintuitive behaviour is due to the larger absolute uncertainty of the background component relative to the difference between the number of data events and background events in the \textit{Medium} WP, which is roughly three times larger than in the \textit{Tight} WP.

The corresponding SF will be applied as an event weight to simulated events containing a di-$\tau$ object geometrically matched to a $\tauhad^{+}\tauhad^{-}$ pair at particle-level, in future ATLAS studies utilising this tagger.
In analyses targeting a different phase space compared with this measurement, additional uncertainties may be required -- likely in the form of an extrapolation uncertainty in one or more kinematic features of the di-$\tau$ object.


%
%
\FloatBarrier

%
\section{Conclusions}
\label{sec:conclusion}
A tagging algorithm for hadronically decaying Lorentz-boosted di-$\tau$ systems originating from decays of low-mass particles with \pt smaller than 300~\GeV~is presented.
The identification algorithm applies a BDT using features of the di-$\tau$ object reconstructed based on tracking and calorimeter information from the ATLAS detector at the Large Hadron Collider.
A measurement of the identification efficiency was performed with $Z\gamma$ events using proton--proton collision data at $\sqrt{s}=13$~\TeV recorded with the ATLAS detector between 2015 and 2018, corresponding to an integrated luminosity of $139\,\mbox{fb}^{-1}$.
Two BDT-based di-$\tau$ identification criteria, \textit{Medium} and \textit{Tight}, were defined, corresponding to estimated signal efficiencies of approximately 70\% and 34\%, with measured background rejection rates of approximately 240 and 3600, respectively. The measured identification scale factors are $1.00\pm0.37$ (tot.) and $1.01\pm0.27$ (tot.), respectively, demonstrating good data-to-simulation agreement for the di-$\tau$ object modelling.
This novel measurement allows using the tagger as an alternative to the standard ATLAS \tauhadvis reconstruction. Physics analyses may
combine both methods to reject fake di-$\tau$ background, improving search sensitivities at \pt values characteristic of the kinematic phase space relevant to light resonance searches.


%
\section*{Acknowledgements}
%

%
%

%
%

We thank CERN for the very successful operation of the LHC and its injectors, as well as the support staff at
CERN and at our institutions worldwide without whom ATLAS could not be operated efficiently.

The crucial computing support from all WLCG partners is acknowledged gratefully, in particular from CERN, the ATLAS Tier-1 facilities at TRIUMF/SFU (Canada), NDGF (Denmark, Norway, Sweden), CC-IN2P3 (France), KIT/GridKA (Germany), INFN-CNAF (Italy), NL-T1 (Netherlands), PIC (Spain), RAL (UK) and BNL (USA), the Tier-2 facilities worldwide and large non-WLCG resource providers. Major contributors of computing resources are listed in Ref.~\cite{ATL-SOFT-PUB-2025-001}.

We gratefully acknowledge the support of ANPCyT, Argentina; YerPhI, Armenia; ARC, Australia; BMWFW and FWF, Austria; ANAS, Azerbaijan; CNPq and FAPESP, Brazil; NSERC, NRC and CFI, Canada; CERN; ANID, Chile; CAS, MOST and NSFC, China; Minciencias, Colombia; MEYS CR, Czech Republic; DNRF and DNSRC, Denmark; IN2P3-CNRS and CEA-DRF/IRFU, France; SRNSFG, Georgia; BMBF, HGF and MPG, Germany; GSRI, Greece; RGC and Hong Kong SAR, China; ISF and Benoziyo Center, Israel; INFN, Italy; MEXT and JSPS, Japan; CNRST, Morocco; NWO, Netherlands; RCN, Norway; MNiSW, Poland; FCT, Portugal; MNE/IFA, Romania; MSTDI, Serbia; MSSR, Slovakia; ARIS and MVZI, Slovenia; DSI/NRF, South Africa; MICIU/AEI, Spain; SRC and Wallenberg Foundation, Sweden; SERI, SNSF and Cantons of Bern and Geneva, Switzerland; NSTC, Taipei; TENMAK, T\"urkiye; STFC/UKRI, United Kingdom; DOE and NSF, United States of America.

Individual groups and members have received support from BCKDF, CANARIE, CRC and DRAC, Canada; CERN-CZ, FORTE and PRIMUS, Czech Republic; COST, ERC, ERDF, Horizon 2020, ICSC-NextGenerationEU and Marie Sk{\l}odowska-Curie Actions, European Union; Investissements d'Avenir Labex, Investissements d'Avenir Idex and ANR, France; DFG and AvH Foundation, Germany; Herakleitos, Thales and Aristeia programmes co-financed by EU-ESF and the Greek NSRF, Greece; BSF-NSF and MINERVA, Israel; NCN and NAWA, Poland; La Caixa Banking Foundation, CERCA Programme Generalitat de Catalunya and PROMETEO and GenT Programmes Generalitat Valenciana, Spain; G\"{o}ran Gustafssons Stiftelse, Sweden; The Royal Society and Leverhulme Trust, United Kingdom.

In addition, individual members wish to acknowledge support from Armenia: Yerevan Physics Institute (FAPERJ); CERN: European Organization for Nuclear Research (CERN PJAS); Chile: Agencia Nacional de Investigaci\'on y Desarrollo (FONDECYT 1230812, FONDECYT 1230987, FONDECYT 1240864); China: Chinese Ministry of Science and Technology (MOST-2023YFA1605700, MOST-2023YFA1609300), National Natural Science Foundation of China (NSFC - 12175119, NSFC 12275265, NSFC-12075060); Czech Republic: Czech Science Foundation (GACR - 24-11373S), Ministry of Education Youth and Sports (FORTE CZ.02.01.01/00/22\_008/0004632), PRIMUS Research Programme (PRIMUS/21/SCI/017); EU: H2020 European Research Council (ERC - 101002463); European Union: European Research Council (ERC - 948254, ERC 101089007), Horizon 2020 Framework Programme (MUCCA - CHIST-ERA-19-XAI-00), European Union, Future Artificial Intelligence Research (FAIR-NextGenerationEU PE00000013), Italian Center for High Performance Computing, Big Data and Quantum Computing (ICSC, NextGenerationEU); France: Agence Nationale de la Recherche (ANR-20-CE31-0013, ANR-21-CE31-0013, ANR-21-CE31-0022, ANR-22-EDIR-0002), Investissements d'Avenir Labex (ANR-11-LABX-0012); Germany: Baden-Württemberg Stiftung (BW Stiftung-Postdoc Eliteprogramme), Deutsche Forschungsgemeinschaft (DFG - 469666862, DFG - CR 312/5-2); Italy: Istituto Nazionale di Fisica Nucleare (ICSC, NextGenerationEU), Ministero dell'Università e della Ricerca (PRIN - 20223N7F8K - PNRR M4.C2.1.1); Japan: Japan Society for the Promotion of Science (JSPS KAKENHI JP22H01227, JSPS KAKENHI JP22H04944, JSPS KAKENHI JP22KK0227, JSPS KAKENHI JP23KK0245); Netherlands: Netherlands Organisation for Scientific Research (NWO Veni 2020 - VI.Veni.202.179); Norway: Research Council of Norway (RCN-314472); Poland: Ministry of Science and Higher Education (IDUB AGH, POB8, D4 no 9722), Polish National Agency for Academic Exchange (PPN/PPO/2020/1/00002/U/00001), Polish National Science Centre (NCN 2021/42/E/ST2/00350, NCN OPUS 2023/51/B/ST2/02507, NCN OPUS nr 2022/47/B/ST2/03059, NCN UMO-2019/34/E/ST2/00393, NCN \& H2020 MSCA 945339, UMO-2020/37/B/ST2/01043, UMO-2021/40/C/ST2/00187, UMO-2022/47/O/ST2/00148, UMO-2023/49/B/ST2/04085, UMO-2023/51/B/ST2/00920); Slovenia: Slovenian Research Agency (ARIS grant J1-3010); Spain: Generalitat Valenciana (Artemisa, FEDER, IDIFEDER/2018/048), Ministry of Science and Innovation (MCIN \& NextGenEU PCI2022-135018-2, MICIN \& FEDER PID2021-125273NB, RYC2019-028510-I, RYC2020-030254-I, RYC2021-031273-I, RYC2022-038164-I); Sweden: Carl Trygger Foundation (Carl Trygger Foundation CTS 22:2312), Swedish Research Council (Swedish Research Council 2023-04654, VR 2018-00482, VR 2021-03651, VR 2022-03845, VR 2022-04683, VR 2023-03403), Knut and Alice Wallenberg Foundation (KAW 2018.0458, KAW 2019.0447, KAW 2022.0358); Switzerland: Swiss National Science Foundation (SNSF - PCEFP2\_194658); United Kingdom: Leverhulme Trust (Leverhulme Trust RPG-2020-004), Royal Society (NIF-R1-231091); United States of America: U.S. Department of Energy (ECA DE-AC02-76SF00515), Neubauer Family Foundation.

%
%


%
%
%
%
%
%
%

%
%
%
%
%
%

%
%
%
\printbibliography
\clearpage
 
\begin{flushleft}
\hypersetup{urlcolor=black}
{\Large The ATLAS Collaboration}

\bigskip

\AtlasOrcid[0000-0002-6665-4934]{G.~Aad}$^\textrm{\scriptsize 105}$,
\AtlasOrcid[0000-0001-7616-1554]{E.~Aakvaag}$^\textrm{\scriptsize 17}$,
\AtlasOrcid[0000-0002-5888-2734]{B.~Abbott}$^\textrm{\scriptsize 124}$,
\AtlasOrcid[0000-0002-0287-5869]{S.~Abdelhameed}$^\textrm{\scriptsize 120a}$,
\AtlasOrcid[0000-0002-1002-1652]{K.~Abeling}$^\textrm{\scriptsize 57}$,
\AtlasOrcid[0000-0001-5763-2760]{N.J.~Abicht}$^\textrm{\scriptsize 51}$,
\AtlasOrcid[0000-0002-8496-9294]{S.H.~Abidi}$^\textrm{\scriptsize 30}$,
\AtlasOrcid[0009-0003-6578-220X]{M.~Aboelela}$^\textrm{\scriptsize 46}$,
\AtlasOrcid[0000-0002-9987-2292]{A.~Aboulhorma}$^\textrm{\scriptsize 36e}$,
\AtlasOrcid[0000-0001-5329-6640]{H.~Abramowicz}$^\textrm{\scriptsize 156}$
% ...
\bigskip
\\

% ...
$^{17}$Department for Physics and Technology, University of Bergen, Bergen; Norway.\\
% ...
$^{30}$Physics Department, Brookhaven National Laboratory, Upton NY; United States of America.\\
% ...
$^{36}$$^{(a)}$Facult\'e des Sciences Ain Chock, Universit\'e Hassan II de Casablanca;$^{(b)}$Facult\'{e} des Sciences, Universit\'{e} Ibn-Tofail, K\'{e}nitra;$^{(c)}$Facult\'e des Sciences Semlalia, Universit\'e Cadi Ayyad, LPHEA-Marrakech;$^{(d)}$LPMR, Facult\'e des Sciences, Universit\'e Mohamed Premier, Oujda;$^{(e)}$Facult\'e des sciences, Universit\'e Mohammed V, Rabat;$^{(f)}$Institute of Applied Physics, Mohammed VI Polytechnic University, Ben Guerir; Morocco.\\
% ...
$^{46}$Physics Department, Southern Methodist University, Dallas TX; United States of America.\\
% ...
$^{51}$Fakult\"{a}t Physik , Technische Universit{\"a}t Dortmund, Dortmund; Germany.\\
% ...
$^{57}$II. Physikalisches Institut, Georg-August-Universit\"{a}t G\"ottingen, G\"ottingen; Germany.\\
% ...
$^{105}$CPPM, Aix-Marseille Universit\'e, CNRS/IN2P3, Marseille; France.\\
% ...
$^{120}$$^{(a)}$New York University Abu Dhabi, Abu Dhabi;$^{(b)}$United Arab Emirates University, Al Ain; United Arab Emirates.\\
% ...
$^{124}$Homer L. Dodge Department of Physics and Astronomy, University of Oklahoma, Norman OK; United States of America.\\
% ...
$^{156}$Raymond and Beverly Sackler School of Physics and Astronomy, Tel Aviv University, Tel Aviv; Israel.\\
% ...
$^{a}$ Also Affiliated with an institute covered by a cooperation agreement with CERN.\\
$^{b}$ Also at An-Najah National University, Nablus; Palestine.\\
$^{c}$ Also at Borough of Manhattan Community College, City University of New York, New York NY; United States of America.\\
$^{d}$ Also at Center for High Energy Physics, Peking University; China.\\
$^{e}$ Also at Center for Interdisciplinary Research and Innovation (CIRI-AUTH), Thessaloniki; Greece.\\
$^{f}$ Also at CERN, Geneva; Switzerland.\\
% ...

\end{flushleft}


%

%
%

%

%
%
%
%

\end{document}